\documentclass[aps,prc,twocolumn,floatfix,superscriptaddress,amsmath,amssymb]{revtex4-2}

\usepackage[utf8]{inputenc} 
\usepackage[T1]{fontenc}
\usepackage{dcolumn}
\usepackage{braket}
\usepackage{float}
\usepackage{graphicx}
\usepackage{microtype}
\usepackage{simplewick}
\usepackage{nicefrac}
\usepackage[colorlinks=true,linkcolor=blue,citecolor=blue,urlcolor=blue]{hyperref}
\usepackage{orcidlink}
\usepackage{slashed}
\usepackage{multirow}
\usepackage[normalem]{ulem}

\graphicspath{{.}}

\begin{document}
    
\title{Saturation of Nuclear Binding from Lattice Hamiltonians}


\author{Maxwell~Rothman\orcidlink{0000-0001-9991-5670}}
\affiliation{Department of Physics and Astronomy, University of
  Tennessee, Knoxville, Tennessee 37996, USA}

\author{Gaute~Hagen\orcidlink{0000-0001-6019-1687}}
\affiliation{Physics Division, Oak Ridge National Laboratory, Oak
  Ridge, Tennessee 37831, USA}
\affiliation{Department of Physics and Astronomy, University of
  Tennessee, Knoxville, Tennessee 37996, USA}

\author{Matthias~Heinz\orcidlink{0000-0002-6363-0056}}
\affiliation{National Center for Computational Sciences, Oak Ridge National Laboratory, Oak Ridge, Tennessee 37831, USA}
\affiliation{Physics Division, Oak Ridge National Laboratory, Oak
  Ridge, Tennessee 37831, USA}

\author{Thomas~Papenbrock\orcidlink{0000-0001-8733-2849}}
\affiliation{Department of Physics and Astronomy, University of Tennessee, Knoxville, Tennessee 37996, USA}
\affiliation{Physics Division, Oak Ridge National Laboratory, Oak Ridge, Tennessee 37831, USA}

\begin{abstract}
There is a conundrum regarding the binding of $\alpha$ particles in nuclei. On one hand, auxiliary-field Monte Carlo simulations of Hamiltonians on discrete spatial lattices proposed that attractive two-nucleon potentials, alone or together with attractive three-nucleon potentials, yield accurate nuclear binding. On the other hand, such Hamiltonians typically overbind all but the lightest nuclei in continuum-space approaches. We address this puzzle by performing Hartree-Fock computations of the light nuclei $^4$He, $^8$Be, $^{12}$C, and $^{16}$O, and of nuclear and neutron matter using established lattice Hamiltonians. These variational upper bounds for the ground-state energies show that the Hamiltonians with only two-nucleon potentials do not yield accurate binding, in contrast to the results from auxiliary-field Monte Carlo simulations. The case is different for Hamiltonians with three-nucleon potentials although it is the dense packing on the lattice -- and not repulsive potentials -- that yield a constant binding energy per nucleon.    
\end{abstract}
\maketitle

\textit{Introduction.---} 
Nuclear Hamiltonians from effective field theories of quantum chromodynamics~\cite{epelbaum2009,machleidt2011,gezerlis2013,piarulli2015,ekstrom2015a,piarulli2016,Hammer:2019poc,reinert2018,maris2021}, the fundamental theory of the strong force, consist of two- and three-body interactions and are rather complicated. There is a scale-dependent balance between attractive and repulsive forces and a complex spin and isospin structure. At next-to-next-to-leading order in the Weinberg power counting, Hamiltonians from chiral effective field theory have about 15 unknown low-energy constants.  While such Hamiltonians have been used to predict~\cite{hagen2015,hagen2017,morris2018,Bonaiti:2025bsb} and understand~\cite{gysbers2019,hu2022,sun2025,ding2026} properties of atomic nuclei, there have also been efforts to understand key elements of nuclear structure using  much simpler Hamiltonians~\cite{konig2016,elhatisari2016,elhatisari2017,lu2019,gnech2024}.

The accurate reproduction of nuclear saturation, i.e., binding energies of about 8~MeV per nucleon and charge radii that scale as $A^{1/3}$ for a nucleus with mass number $A$ is a key test of nuclear Hamiltonians. This implies that the equation of state for nuclear matter with equal parts neutrons and protons has a minimum at a density of about $0.16$~fm$^{-3}$ (which is approximately the central density of stable medium-mass nuclei) and at an energy of about $-16$~MeV per nucleon.
Accurate nuclear saturation is essential for reproducing bulk properties of nuclei, and for this reason many nuclear Hamiltonians are optimized to nuclear matter properties or bulk properties of medium-mass nuclei~\cite{ekstrom2015a, drischler2019, jiang2020, arthuis2024, elhatisari2024}.

In this work we focus on interactions from effective field theories of quantum chromodynamics that are formulated on discrete lattices. At leading order of chiral effective field theory~\cite{epelbaum2009,machleidt2011} the two-body potential consists of the one-pion exchange and two $s$-wave contacts. There are no three-nucleon forces at this order. Specific formulations are presented in Refs.~\cite{elhatisari2016,elhatisari2017}. At leading order of pion-less effective field theory, the potential consists of two-body $s$-wave contacts and a three-body contact~\cite{bedaque2002}. A specific formulation was given in Ref.~\cite{lu2019}. These lattice formulations use two- and three-body contacts that are spin/isospin symmetric and consist of products of densities with nonlocal and local smearing; details are presented below.       

Auxiliary-field Monte Carlo simulations of these lattice Hamiltonians found (i) that simple two-body interactions  can yield accurate binding energies and radii for nuclei up to oxygen~\cite{elhatisari2017}, (ii) that nuclear binding and $\alpha$-particle clustering is sensitive to the mixture of local and nonlocal smearing in the contact interactions~\cite{elhatisari2016}, and (iii) that attractive two- and three-body potentials yield essential elements of nuclear binding~\cite{lu2019}.

These lattice results are somewhat surprising because the two-body lattice interactions of Refs.~\cite{elhatisari2016,elhatisari2017} are very soft (the lattice spacing of $a=1.97$~fm corresponds to a momentum cutoff of $\pi/a\approx 314$~MeV) and overall attractive in nature. The quantitative description of nuclei using Hamiltonians with only attractive two-body forces is at odds with computations that use Hamiltonians formulated in contiuum space. There, interactions from chiral effective field theory at leading order do not bind $\alpha$-particles into nuclei~\cite{maris2021} (at least at momentum cutoffs of 450 and 500~MeV), and phase-shift equivalent two-body Hamiltonians alone generally fail to saturate at accurate densities~\cite{hebeler2011} or overbind medium-mass nuclei~\cite{hagen2007b,ekstrom2013}. Three-nucleon interactions are critical for an accurate description of nuclear structure~\cite{Carlson1983NPA_Urbana3NF, pieper2001b, hammer2013, hebeler2015, hebeler2021}, and this is particularly so for nucleon-nucleon interactions with a low momentum cutoff~\cite{hagen2007b,hebeler2011}. It is the purpose of this Letter to resolve this conundrum. 

We note that the lattice computations of Refs.~\cite{elhatisari2016,elhatisari2017,lu2019} escaped scrutiny so far because alternative ab initio methods~\cite{dickhoff2004,barrett2013,soma2013,hagen2014,hergert2016,launey2016,stroberg2019,tichai2020,heinz2021} employ the harmonic oscillator basis or work in the continuum~\cite{piarulli2017}. However, the recent publication of \texttt{NuLattice}~\cite{Rothman:2025uza}, a publicly available Python package for ab initio computations on lattices, allows us now to compare with auxiliary-field Monte Carlo simulations of lattice Hamiltonians.

\textit{Lattice Hamiltonians.---} 
We employ a discrete three-dimensional spatial lattice as the single-particle basis and use Hamiltonians from Refs.~\cite{elhatisari2016,elhatisari2017,lu2019}. The papers~\cite{elhatisari2016,elhatisari2017} used a lattice spacing of $1.97$~fm and Hamiltonians from chiral effective field theory at leading order. These consist of the kinetic energy $T$, the one-pion exchange term $V_{\text{OPE}}$, and a short-range term $V_0$. In Ref.~\cite{elhatisari2017} the Hamiltonian is 
\begin{equation}
\label{ham_chiral}
	H=T+V_{\text{OPE}}+V_0 \ .	
\end{equation}
Expressions for the kinetic energy and the one-pion exchange are given in \hyperref[sec:end_matter]{End Matter}. Here we discuss the short-range potential $V_0$ in more detail and follow Refs.~\cite{elhatisari2016,elhatisari2017}. 
Using a combined spin and isospin index $i$, one defines a smeared annihilation operator on the lattice site $\mathbf{n}$ as
\begin{equation}
\label{tilde-a}
	\tilde{a}_{i}(\mathbf{n}) = a_{i}(\mathbf{n})+s_{NL}\sum\limits_{|\mathbf{n}-\mathbf{n'}|=1}a_{i}(\mathbf{n'})\,.
\end{equation} 
One then defines the smeared density operator as
\begin{equation}
\label{tilderho}
	\tilde{\rho}(\mathbf{n}) = \sum\limits_{i}\left[\tilde{a}^\dagger_{i}(\mathbf{n})\tilde{a}_{i}(\mathbf{n}) + s_L\sum\limits_{|\mathbf{n}-\mathbf{n'}|=1}\tilde{a}^\dagger_{i}(\mathbf{n'})\tilde{a}_{i}(\mathbf{n'})\right]
    .
\end{equation}
The short-range potential of Ref.~\cite{elhatisari2017} is then simply
\begin{equation}
\label{cont}
	V_0 = \frac{c_0}{2}\sum\limits_n :\tilde{\rho}(\mathbf{n})^2:\,.
\end{equation}
Here, the colons indicate normal ordering. We note that the potential is nonlocal, i.e., it does not act multiplicative, when $s_{NL}\ne 0$. The parameters are $c_0 = -0.185$, and $s_{NL} = s_L= 0.08$, see Ref.~\cite{elhatisari2017}. The potential~(\ref{cont}) is in lattice units and needs to be multiplied with $\hbar/a=100$~MeV in actual computations. 

The Hamiltonian  of Ref.~\cite{elhatisari2016} also has the form~(\ref{ham_chiral}) but employs different short-range two-body potentials. We see from Eq.~\eqref{tilderho} that $\tilde{\rho}(\mathbf{n}) = \tilde{\rho}(s_{NL},s_L;\mathbf{n})$. Thus one can consider potentials made from local densities (where $s_{NL}=0$) and from nonlocal densities (where $s_{L}=0$). Furthermore, one can insert spin and isospin operators acting on these densities. The upshot is that Ref.~\cite{elhatisari2016} proposed a Hamiltonian 
\begin{equation}
\label{hamB}
    H_B = T+V_{\text{OPE}}+V_{L} +V_{NL}\,.	
\end{equation}
where $V_{L}$ and $V_{NL}$ denote local and nonlocal short-range potentials, respectively, that also contain spin-isospin operators. A key result of that work was that a purely nonlocal interaction fails to bind $\alpha$ particles into nuclei while the Hamiltonian $H_B$ yields accurate results for light nuclei. Details of the Hamiltonian $H_B$ are presented in \hyperref[sec:end_matter]{End Matter}. 

Finally, we follow Ref.~\cite{lu2019} and use a Hamiltonian from pion-less effective field theory. It has the form  
    \begin{equation}
    \label{nopi}
    H_{\slashed{\pi}}=T+V_0 + W\,.
\end{equation}
Here,
\begin{equation}
\label{W}
    W \equiv \frac{c_3}{3!}\sum_{\mathbf{n}}: \tilde{\rho}(\mathbf{n})^3: 
\end{equation}
is the three-body potential. For this Hamiltonian, the lattice spacing is $a=1.32$~fm (corresponding to  a momentum cutoff of $\pi/a\approx 471$~MeV), and the constants of the potential are $s_{NL}=0.5$, $s_L=0.061$, $c_0=-3.41\times 10^{-7}$~MeV$^{-2}$, and $c_3=-1.4\times 10^{-14}$~MeV$^{-5}$. (In numerical computations, the last two constants need to be multiplied with three and six powers of $\hbar/a=150$~MeV, respectively.) We note that the two- and three-body potentials are both attractive and that the nonlocal smearing $s_{NL}$ is about an order of magnitude larger than what was used in Refs.~\cite{elhatisari2016,elhatisari2017}.
We also note already at this point that the auxiliary-field Monte Carlo simulations of Ref.~\cite{lu2019} used a temporal lattice spacing (i.e., a time step) of $a_t=1/(1000~{\rm MeV})$ in the computations with the Hamiltonian~(\ref{nopi}) while the Hamiltonians~(\ref{ham_chiral}) and (\ref{hamB}) of Refs.~\cite{elhatisari2016} and \cite{elhatisari2017}, respectively, were solved with a much larger  $a_t=1/(150~{\rm MeV})$.  

\textit{Finite nuclei.---}
We start with Hartree-Fock computations of finite nuclei based on the Hamiltonian~(\ref{ham_chiral}) from chiral effective field theory at leading order, using the parameterizations and lattice ($L=6$) of Ref.~\cite{elhatisari2017}.
Our computations start from simple localized initial states (i.e., compact clusters of four nucleons on neighboring sites) and solve the Hartree-Fock equations self consistently.
Our initial states are too compact, and 
the Hartree-Fock iterations relax the density. At fixed $L$ the resulting Hartree-Fock energies are variational upper bounds on the exact ground-state energy,
which would require additional correlations beyond the mean-field level to compute.

\begin{table}[tb!]
\renewcommand{\arraystretch}{1.2}
\centering
\caption{Variational upper bounds from Hartree-Fock (HF) for ground-state energies (in MeV) for various nuclei and the Hamiltonians $H$ [from Eq.~(\ref{ham_chiral})] and $H_B$ [from Eq.~(\ref{hamB})] compared to the results from auxiliary-field Monte Carlo (AFMC) simulations~\cite{elhatisari2017} and \cite{elhatisari2016} that used the same Hamiltonians, respectively, and experiment~\cite{wang2021}.}
\label{tab:res_combo}
\renewcommand{\arraystretch}{1.3}
\begin{ruledtabular}
\begin{tabular}{c|dd|dd|r}
\multirow{2}{*}{Nucl.} & \multicolumn{2}{c|}{$H$ from Ref.~\cite{elhatisari2017}} & \multicolumn{2}{c|}{$H_B$ from Ref.~\cite{elhatisari2016}}& 
\multirow{2}{*}{Exp.}\\
 & \multicolumn{1}{c}{HF} & \multicolumn{1}{c|}{AFMC} & \multicolumn{1}{c}{HF} & \multicolumn{1}{c|}{AFMC} & \\
\hline
$^{4}$He &   -19.81 & -25.4   &  -32.19  &  -29.2  &  $-28.3$\\
$^{8}$Be &   -63.17 & -51.9   &  -76.37  &  -59.7  &  $-56.5$\\
$^{12}$C &  -137.28 & -83.8   &  -139.83 &  -95.0  &  $-92.2$\\
$^{16}$O &  -213.37 & -128.2  &  -222.06 & -135.4  & $-127.6$\\
\end{tabular}
\end{ruledtabular}
\end{table}

Our results for the Hamiltonian $H$ of Ref.~\cite{elhatisari2017} and Eq.~(\ref{ham_chiral}) are summarized in Table~\ref{tab:res_combo} and compared to  auxiliary-field Monte Carlo simulations of the same publication.
The Hartree-Fock energy of $^4$He is a plausible upper bound.
For heavier nuclei, however,
the variational bounds from Hartree Fock are significantly lower than the results from auxiliary-field Monte Carlo~\cite{elhatisari2017}.
The nuclei $^{8}$Be, $^{12}$C, and $^{16}$O are overbound with 7.9, 11.4, and 13.3~MeV binding energy per particle, respectively.
We see no sign of saturation, and we are far from the nuclear binding energies of about 8~MeV per nucleon one observes in nature.

We repeat the Hartree-Fock computations for the Hamiltonian $H_B$ of Eq.~(\ref{hamB}) with the parameterization and lattice  ($L=6$) of Ref.~\cite{elhatisari2016}. Results are also shown in Table~\ref{tab:res_combo} and compared to the auxiliary-field Monte Carlo simulations of that publication. The latter are all above the variational upper bounds from Hartree Fock.
The results of Table~\ref{tab:res_combo}  show that the auxiliary-field Monte Carlo simulations of Refs.~\cite{elhatisari2016,elhatisari2017}, which tuned the Hamiltonians to reproduce experiment, do not accurately solve the Hamiltonians presented in those papers~\footnote{
The auxiliary field Monte Carlo calculations of Refs.~\cite{epelbaum2010,epelbaum2010b,epelbaum2011,epelbaum2012,epelbaum2013,epelbaum2014,elhatisari2016,elhatisari2017} solved the many-body problem by applying the transfer matrix formalism~\cite{lee2009} with a discrete temporal lattice spacing $a_t=1/(150\,\mathrm{MeV})$.
Such calculations are not equivalent to calculations with $a_t\to0$ (see Supplemental Material).
Based on our comparison to Refs.~\cite{elhatisari2016,elhatisari2017},
the differences are significant. 
} 
and that the corresponding Hamiltonians do not produce accurate nuclear saturation.

\textit{Nuclear matter.---} 
For our nuclear matter computations we consider lattices of different extent $L$. The kinetic energy is a one-body operator and can be represented by a matrix of dimension $D=4L^3$. In a first step we compute the eigenvalues $T_k$ and eigenstates $|T_k\rangle$ with $k=1,\ldots,D$ of the kinetic energy and store them in order of increasing eigenvalues. The eigenstates are translationally invariant and can therefore serve in computations of homogeneous nuclear matter. 

The eigenvalues $T_k$ come in sets of degenerate numbers that reflect the cubic symmetry of the lattice. On sufficiently large lattices, the ordered eigenvalues change after eigenstate numbers 4, 28, 76, 108, 132, and so on~\cite{gandolfi2009,hagen2013b,marino2024}. In nuclear physics parlance, these are closed-shell configurations on a cubic lattice. We then consider nucleon numbers $A$ that correspond to these closed-shell systems and construct the one-body density matrix
\begin{equation}
    \hat{n} = \sum_{k=1}^A |T_k\rangle \langle T_k|\,.
\end{equation}
Its matrix elements in the spatial lattice basis are $n_q^p=\langle p|\hat{n}|q\rangle$.  
The energy expectation value for the $A$ fermion state with this density matrix is 
\begin{equation}
    E = \sum_{pq}\varepsilon^p_qn_p^q + {1\over 2}\sum_{pqrs}V^{pr}_{qs}n_p^q n^s_r
    +\frac{1}{6}\sum_{\substack{pqr\\suv}}W_{suv}^{pqr}n_p^s n_q^u n_r^v
    \,,  
\end{equation}
where $\varepsilon^p_q$,  $V^{pr}_{qs}$, and $W^{pqr}_{suv}$ are the matrix elements of the one-, two-, and three-body terms in the Hamiltonian (if the latter are present).  These closed-shell configurations are solutions of the Hartree-Fock equations. 

\begin{figure}[t!]
    \centering
    \includegraphics[width=0.49\textwidth]{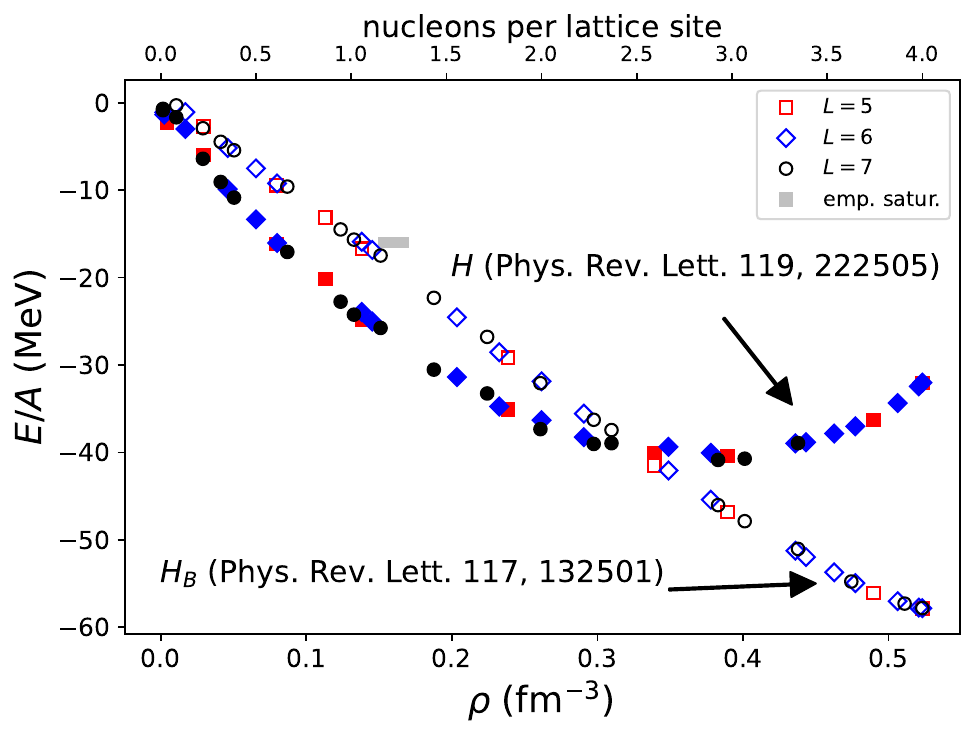}
        \caption{Energy per nucleon ($E/A$) in symmetric nuclear matter as a function of density $\rho$ from mean-field expectation values computed on lattices with extent $L=5$, 6, and 7 with the Hamiltonians of Ref.~\cite{elhatisari2017} (full symbols) and of Ref.~\cite{elhatisari2016} (open symbols). The region ($-16.5$~MeV~$\le E/A\le$$-15.5$~MeV and $0.15$~fm$^{-3}$$\le \rho\le$~$0.17$~fm$^{-3}$) around the empirical saturation point is shown as a gray rectangle. The top $x$ axis shows the density in nucleons per lattice site.} 
    \label{fig:EOS}
\end{figure}

Figure~\ref{fig:EOS} shows the energy per nucleon as a function of the density $\rho=A/(La)^3$ for closed-shell configurations on lattices with extent $L=5$, 6, and 7. Results are shown for the Hamiltonian $H$ of Eq.~(\ref{ham_chiral}) and $H_B$ of Eq.~(\ref{hamB}). We see that the data points from different $L$ fall onto relatively smooth curves for each Hamiltonian. For the Hamiltonian $H$ nuclear matter saturates at a much higher density and binding energy than the empirical saturation point, while  Hamiltonian $H_B$  does not saturate, reaching its lowest energy when the lattice is fully occupied. The top $x$ axis marks the number of nucleons per lattice site.
The equations of state in Fig.~\ref{fig:EOS} indicate that finite nuclei will be overbound, and this is consistent with the results shown in Table~\ref{tab:res_combo}.

\textit{``Why you should trust us~\cite{wirecutter}.''---}
First,  we benchmarked the two-nucleon interaction of the Hamiltonian~(\ref{ham_chiral}) in two-nucleon systems~\cite{elhatisari2026}.
Second, we benchmarked the \texttt{NuLattice} Hartree-Fock code with a modification of the Hartree-Fock workhorse that is used to compute reference states for coupled-cluster computations in the harmonic-oscillator basis~\cite{hagen2014}. 
Third, \texttt{NuLattice} is publicly available~\cite{Rothman:2025uza,Rothman2026NuLatticev1.2}, and the computations that produce the results of this Letter can be replicated and verified.
Fourth, neglecting the small contribution of the one-pion exchange, the analytical and numerical Hartree-Fock energies in a fully occupied lattice agree with each other for the Hamiltonians of this Letter.

\textit{Saturation on lattices.---} We finally turn to Ref.~\cite{lu2019}, where auxiliary-field Monte Carlo computations seem accurate (see below) and examine how attractive two-body and three-body potentials can yield reasonable saturation on lattices. (Further examples are the most recent lattice Hamiltonians of Refs.~\cite{wang2026,agar2026} which also employ attractive three-nucleon forces.)  This is in contrast to continuum formulations where, for soft nuclear interactions, repulsive three-nucleon forces yield saturation~\cite{nogga2004,bogner2005,bogner2010,hebeler2011}. The basic mechanism is simple: In the Thomas-Fermi approximation the kinetic energy scales as $\rho^{5/3}$, while short-range two-body and three-body potential energies scale as $\rho^2$ and $\rho^3$, respectively. Thus, an attractive two-body potential alone overwhelms the kinetic energy at sufficiently high density $\rho$ and leads to a collapse of the system. A repulsive three-body force stabilizes the system.

We perform mean-field computations of neutron matter and present results in Table~\ref{tab:neutrons}. The number of three-body matrix elements $W_{suv}^{pqr}$ is about 1~GB per lattice site for the potential~(\ref{W}), which is prohibitively large. However, as nuclear matter is homogeneous, we only need to compute the expectation value of the three-nucleon potential for a single lattice site (and multiply the result with $L^3$). 

Our results for neutron matter are close to (and above) those from auxiliary-field Monte Carlo simulations~\cite{lu2019}. Assuming that the latter are accurate, we infer that the correlation energy, i.e., the difference between the ground-state energy and the Hartree-Fock energy, is relatively small (about 0.2 MeV per neutron) at $\rho=0.13$~fm$^{-3}$ for an interaction with a momentum cutoff of 471~MeV. While it is not easy to compare potentials with different regulators and cutoffs, we note that  the correlation energies by~\textcite{alp2025} at $\rho=0.13$~fm$^{-3}$ are 0.5~MeV per neutron for the very soft interaction {1.8/2.0~(EM)} of Ref.~\cite{hebeler2011} (whose cutoffs are about 360 and 400~MeV for the nucleon-nucleon and the three-nucleon interaction, respectively) and 1.7~MeV per neutron for the harder interaction $\Delta$NNLO$_{\rm GO}(450)$ of Ref.~\cite{jiang2020} (whose cutoff is 450~MeV).

\begin{table}[t!]
\renewcommand{\arraystretch}{1.2}
\centering
\caption{Energy per particle $E/N$ (in MeV) of neutron matter at a density $\rho$ (in fm$^{-3}$) computed with Hartree Fock using lattices of extent $L$ and number of neutrons $N$, and compared to the auxiliary-field Monte Carlo simulations~\cite{lu2019}.}
\label{tab:neutrons}
\begin{ruledtabular}
\begin{tabular}{ddddr}
L & N & \multicolumn{1}{c}{$\rho$} & \multicolumn{1}{c}{$E/N$} & \multicolumn{1}{c}{$E/N$ Ref.~\cite{lu2019}}\\
\hline
7 & 14 & 0.018 &  3.8 &  2.7 \\
7 & 38 & 0.048 &  5.6 &  5.1 \\
7 & 54 & 0.068 &  7.5 &  7.1 \\
6 & 38 & 0.076 &  8.0 &  7.6 \\
7 & 66 & 0.084 &  9.3 &  8.9 \\
6 & 54 & 0.11  & 11.6 & 11.3 \\
5 & 38 & 0.13  & 13.6 & 13.4 \\
6 & 66 & 0.13  & 14.0 & 13.8 \\
\end{tabular}
\end{ruledtabular}
\end{table}  

\begin{figure}[t!]
    \centering
    \includegraphics[width=0.49\textwidth]{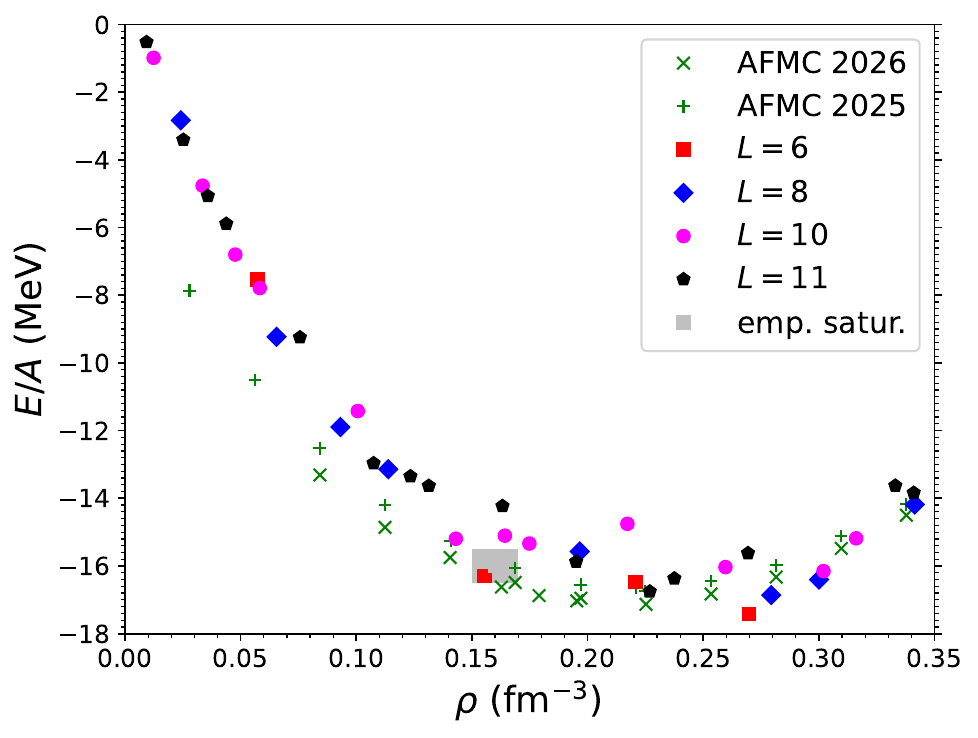}
        \caption{Energy per nucleon $E/A$ in symmetric nuclear matter as a function of density $\rho$ from mean-field expectation values computed on lattices with extent $L$ using the Hamiltonian of Ref.~\cite{lu2019} [See Eq.~(\ref{nopi})]. The  empirical saturation point is shown as a gray rectangle. The auxiliary-field Monte Carlo results ``AFMC~2025'' and ``AFMC~2026'' are  from Refs.~\cite{niu2025} and \cite{agar2026}, respectively.} 
    \label{fig:EOS_2018}
\end{figure}

Mean-field results for the equation of state of symmetric nuclear matter are shown in Fig.~\ref{fig:EOS_2018}. We see that the saturation point is close to the empirical one already at the Hartree-Fock level and that shell effects (which cause the scatter) are significant. For comparison we also show the results from the auxiliary-field Monte Carlo, labeled as ``AFMC 2025'' and  ``AFMC 2026'', from  Refs.~\cite{niu2025} and \cite{agar2026}, respectively. (These differ slightly from each other because of different lattice extents $L$ and infinite-time extrapolations.) We note that the Hartree-Fock results are only variational upper bounds for computations with identical $L$ and $A$. Apparently, the correlation energy is small.
For a comparison, the correlation energies from Ref. \cite{alp2025} (at the saturation density $\rho=0.16$~fm$^{-3}$) are 3.9~MeV per nucleon for the interaction {1.8/2.0~(EM)} and 6.7~MeV per nucleon for  $\Delta$NNLO$_{\rm GO}(450)$.

\begin{figure}[t!]
    \centering
    \includegraphics[width=0.49\textwidth]{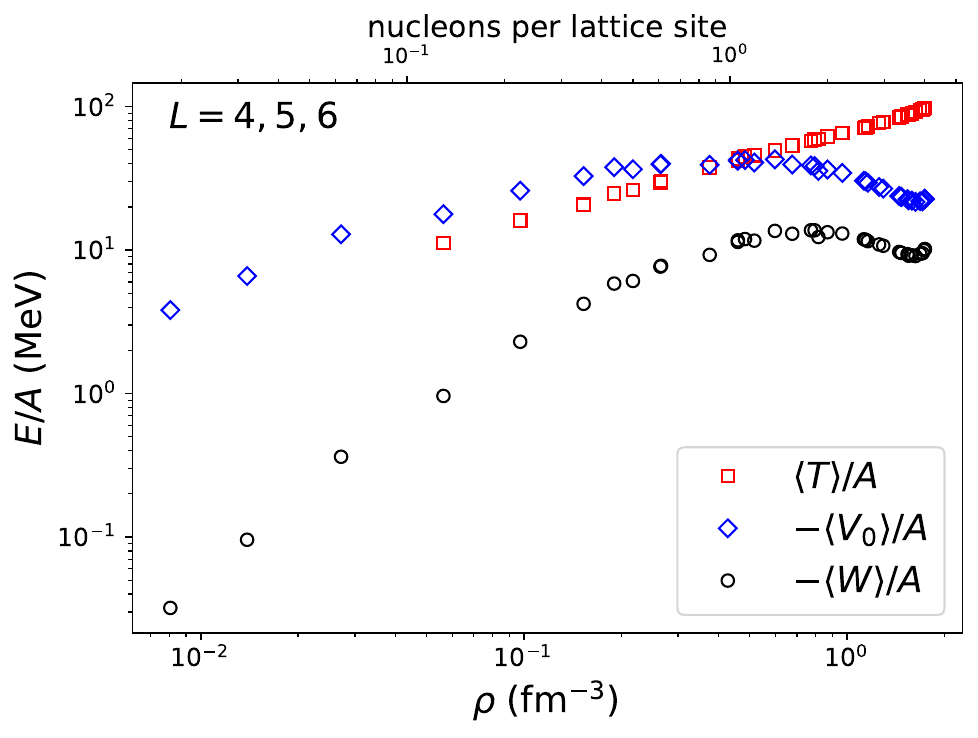}
        \caption{Positive expectation values of the kinetic energy $T$, the two-body potential $(-V_0)$ and the three-body interaction $(-W)$ energy per nucleon ($E/A$) in symmetric nuclear matter as a function of density $\rho$ from mean-field expectation values computed on lattices with extent $L=4$, 5, and 6 using the Hamiltonian of Ref.~\cite{lu2019}.
        The top $x$ axis shows the density in nucleons per lattice site.} 
    \label{fig:EOS-contrib}
\end{figure}

To understand the saturation mechanism we computed the expectation values of the kinetic energy per nucleon $\langle T\rangle/A$ and the positive expectation values $-\langle V_0\rangle/A$ and $-\langle W\rangle/A$ on lattices with $L=4$, 5, 6 and show the results in Fig.~\ref{fig:EOS-contrib}. (The kinetic energy vanishes for the lowest density in each lattice system and is not shown on the log plot.) The kinetic energy per nucleon is approximately proportional to $\rho^{2/3}$ for sufficiently large densities, as expected from Thomas-Fermi theory~\cite{lieb1976}. At low densities, the magnitude of the attractive two- and three-body potentials increase as $\rho$ and $\rho^2$, respectively, and this is much faster than the kinetic energy. However, the potential energies peak at densities of about one nucleon per lattice site and then decrease in magnitude. This is due to effects of the finite lattice on the nonlocal interaction. We recall that a nonlocal potential is not multiplicative on the lattice and really moves nucleons to neighboring lattice sites. 
As the lattice becomes densely packed, Pauli blocking prevents nucleons from moving, reducing the attraction from the interaction.
Apparently, this transition happens when a density of one nucleon per lattice site is reached.
Thus, saturation is obtained from the dominance of kinetic over potential energies due to a dense packing of nucleons. It is a lattice artifact.

\textit{Summary and discussion.---}
We scrutinized the claims of Refs.~\cite{elhatisari2016,elhatisari2017,lu2019} that simple lattice Hamiltionians with purely attractive potentials yield nuclear saturation. Our Hartree-Fock computations of the finite nuclei and of symmetric nuclear matter showed that the Hamiltonians from Refs.~\cite{elhatisari2016,elhatisari2017} do not saturate properly and that the auxiliary field Monte Carlo simulations reported in those works did not accurately solve the corresponding Hamiltonians. We also showed that the attractive two-body and three-nucleon potentials of Ref.~\cite{lu2019} approximately yield a physical saturation point at the Hartree-Fock level. This saturation comes from the dominance of the kinetic energy over potential energies once occupations become saturated on the lattice; it is a lattice artifact.
With view on Refs.~\cite{elhatisari2016,elhatisari2017} we conclude that an ab initio understanding of binding and clustering of $\alpha$ particles in atomic nuclei~\cite{freer2018,Otsuka:2026vwf} remains an open challenge. 

\acknowledgments We are grateful to Serdar Elhatisari  for benchmarks of two-nucleon systems and many helpful discussions. We also thank Dick Furnstahl, Dean Lee, Witek Nazarewicz, and Achim Schwenk for useful discussions. We thank participants of the INT Program ``Nuclear Hamiltonians for Advancing Nuclear Physics and Beyond'' for questions about the continuum limit. This work was supported by the U.S.~Department of Energy, Office of Science, Office of Nuclear Physics, under award Nos.~DE-FG02-96ER40963 and DE-SC0026198; by the U.S.~Department of Energy, Office of Science, Office of Advanced Scientific Computing Research and Office of Nuclear Physics, Scientific Discovery through Advanced Computing (SciDAC) program (SciDAC-5 NUCLEI); and by the Laboratory Directed Research and Development Program of Oak Ridge National Laboratory, managed by UT-Battelle, LLC, for the U.S.\ Department of Energy. This research used resources of the Oak Ridge Leadership Computing Facility at the Oak Ridge National Laboratory, which is supported by the Advanced Scientific Computing Research programs in the Office of Science of the U.S.~Department of Energy under Contract No.~DE-AC05-00OR22725. 

\section*{Data Availability}
The data supporting this study may be reproduced using the published update to \texttt{NuLattice}~\cite{Rothman2026NuLatticev1.2}.

\bibliography{master}

\begin{thebibliography}{73}%
\makeatletter
\providecommand \@ifxundefined [1]{%
 \@ifx{#1\undefined}
}%
\providecommand \@ifnum [1]{%
 \ifnum #1\expandafter \@firstoftwo
 \else \expandafter \@secondoftwo
 \fi
}%
\providecommand \@ifx [1]{%
 \ifx #1\expandafter \@firstoftwo
 \else \expandafter \@secondoftwo
 \fi
}%
\providecommand \natexlab [1]{#1}%
\providecommand \enquote  [1]{``#1''}%
\providecommand \bibnamefont  [1]{#1}%
\providecommand \bibfnamefont [1]{#1}%
\providecommand \citenamefont [1]{#1}%
\providecommand \href@noop [0]{\@secondoftwo}%
\providecommand \href [0]{\begingroup \@sanitize@url \@href}%
\providecommand \@href[1]{\@@startlink{#1}\@@href}%
\providecommand \@@href[1]{\endgroup#1\@@endlink}%
\providecommand \@sanitize@url [0]{\catcode `\\12\catcode `\$12\catcode `\&12\catcode `\#12\catcode `\^12\catcode `\_12\catcode `\%12\relax}%
\providecommand \@@startlink[1]{}%
\providecommand \@@endlink[0]{}%
\providecommand \url  [0]{\begingroup\@sanitize@url \@url }%
\providecommand \@url [1]{\endgroup\@href {#1}{\urlprefix }}%
\providecommand \urlprefix  [0]{URL }%
\providecommand \Eprint [0]{\href }%
\providecommand \doibase [0]{https://doi.org/}%
\providecommand \selectlanguage [0]{\@gobble}%
\providecommand \bibinfo  [0]{\@secondoftwo}%
\providecommand \bibfield  [0]{\@secondoftwo}%
\providecommand \translation [1]{[#1]}%
\providecommand \BibitemOpen [0]{}%
\providecommand \bibitemStop [0]{}%
\providecommand \bibitemNoStop [0]{.\EOS\space}%
\providecommand \EOS [0]{\spacefactor3000\relax}%
\providecommand \BibitemShut  [1]{\csname bibitem#1\endcsname}%
\let\auto@bib@innerbib\@empty
\bibitem [{\citenamefont {Epelbaum}\ \emph {et~al.}(2009)\citenamefont {Epelbaum}, \citenamefont {Hammer},\ and\ \citenamefont {Mei\ss{}ner}}]{epelbaum2009}%
  \BibitemOpen
  \bibfield  {author} {\bibinfo {author} {\bibfnamefont {E.}~\bibnamefont {Epelbaum}}, \bibinfo {author} {\bibfnamefont {H.-W.}\ \bibnamefont {Hammer}},\ and\ \bibinfo {author} {\bibfnamefont {U.-G.}\ \bibnamefont {Mei\ss{}ner}},\ }\bibfield  {title} {\bibinfo {title} {Modern theory of nuclear forces},\ }\href {https://doi.org/10.1103/RevModPhys.81.1773} {\bibfield  {journal} {\bibinfo  {journal} {Rev. Mod. Phys.}\ }\textbf {\bibinfo {volume} {81}},\ \bibinfo {pages} {1773} (\bibinfo {year} {2009})}\BibitemShut {NoStop}%
\bibitem [{\citenamefont {Machleidt}\ and\ \citenamefont {Entem}(2011)}]{machleidt2011}%
  \BibitemOpen
  \bibfield  {author} {\bibinfo {author} {\bibfnamefont {R.}~\bibnamefont {Machleidt}}\ and\ \bibinfo {author} {\bibfnamefont {D.}~\bibnamefont {Entem}},\ }\bibfield  {title} {\bibinfo {title} {Chiral effective field theory and nuclear forces},\ }\href {https://doi.org/10.1016/j.physrep.2011.02.001} {\bibfield  {journal} {\bibinfo  {journal} {Phys. Rep.}\ }\textbf {\bibinfo {volume} {503}},\ \bibinfo {pages} {1 } (\bibinfo {year} {2011})}\BibitemShut {NoStop}%
\bibitem [{\citenamefont {Gezerlis}\ \emph {et~al.}(2013)\citenamefont {Gezerlis}, \citenamefont {Tews}, \citenamefont {Epelbaum}, \citenamefont {Gandolfi}, \citenamefont {Hebeler}, \citenamefont {Nogga},\ and\ \citenamefont {Schwenk}}]{gezerlis2013}%
  \BibitemOpen
  \bibfield  {author} {\bibinfo {author} {\bibfnamefont {A.}~\bibnamefont {Gezerlis}}, \bibinfo {author} {\bibfnamefont {I.}~\bibnamefont {Tews}}, \bibinfo {author} {\bibfnamefont {E.}~\bibnamefont {Epelbaum}}, \bibinfo {author} {\bibfnamefont {S.}~\bibnamefont {Gandolfi}}, \bibinfo {author} {\bibfnamefont {K.}~\bibnamefont {Hebeler}}, \bibinfo {author} {\bibfnamefont {A.}~\bibnamefont {Nogga}},\ and\ \bibinfo {author} {\bibfnamefont {A.}~\bibnamefont {Schwenk}},\ }\bibfield  {title} {\bibinfo {title} {{Quantum Monte Carlo} calculations with chiral effective field theory interactions},\ }\href {https://doi.org/10.1103/PhysRevLett.111.032501} {\bibfield  {journal} {\bibinfo  {journal} {Phys. Rev. Lett.}\ }\textbf {\bibinfo {volume} {111}},\ \bibinfo {pages} {032501} (\bibinfo {year} {2013})}\BibitemShut {NoStop}%
\bibitem [{\citenamefont {Piarulli}\ \emph {et~al.}(2015)\citenamefont {Piarulli}, \citenamefont {Girlanda}, \citenamefont {Schiavilla}, \citenamefont {P\'erez}, \citenamefont {Amaro},\ and\ \citenamefont {Arriola}}]{piarulli2015}%
  \BibitemOpen
  \bibfield  {author} {\bibinfo {author} {\bibfnamefont {M.}~\bibnamefont {Piarulli}}, \bibinfo {author} {\bibfnamefont {L.}~\bibnamefont {Girlanda}}, \bibinfo {author} {\bibfnamefont {R.}~\bibnamefont {Schiavilla}}, \bibinfo {author} {\bibfnamefont {R.~N.}\ \bibnamefont {P\'erez}}, \bibinfo {author} {\bibfnamefont {J.~E.}\ \bibnamefont {Amaro}},\ and\ \bibinfo {author} {\bibfnamefont {E.~R.}\ \bibnamefont {Arriola}},\ }\bibfield  {title} {\bibinfo {title} {Minimally nonlocal nucleon-nucleon potentials with chiral two-pion exchange including $\ensuremath{\Delta}$ resonances},\ }\href {https://doi.org/10.1103/PhysRevC.91.024003} {\bibfield  {journal} {\bibinfo  {journal} {Phys. Rev. C}\ }\textbf {\bibinfo {volume} {91}},\ \bibinfo {pages} {024003} (\bibinfo {year} {2015})}\BibitemShut {NoStop}%
\bibitem [{\citenamefont {Ekstr\"om}\ \emph {et~al.}(2015)\citenamefont {Ekstr\"om}, \citenamefont {Jansen}, \citenamefont {Wendt}, \citenamefont {Hagen}, \citenamefont {Papenbrock}, \citenamefont {Carlsson}, \citenamefont {Forss\'en}, \citenamefont {Hjorth-Jensen}, \citenamefont {Navr\'atil},\ and\ \citenamefont {Nazarewicz}}]{ekstrom2015a}%
  \BibitemOpen
  \bibfield  {author} {\bibinfo {author} {\bibfnamefont {A.}~\bibnamefont {Ekstr\"om}}, \bibinfo {author} {\bibfnamefont {G.~R.}\ \bibnamefont {Jansen}}, \bibinfo {author} {\bibfnamefont {K.~A.}\ \bibnamefont {Wendt}}, \bibinfo {author} {\bibfnamefont {G.}~\bibnamefont {Hagen}}, \bibinfo {author} {\bibfnamefont {T.}~\bibnamefont {Papenbrock}}, \bibinfo {author} {\bibfnamefont {B.~D.}\ \bibnamefont {Carlsson}}, \bibinfo {author} {\bibfnamefont {C.}~\bibnamefont {Forss\'en}}, \bibinfo {author} {\bibfnamefont {M.}~\bibnamefont {Hjorth-Jensen}}, \bibinfo {author} {\bibfnamefont {P.}~\bibnamefont {Navr\'atil}},\ and\ \bibinfo {author} {\bibfnamefont {W.}~\bibnamefont {Nazarewicz}},\ }\bibfield  {title} {\bibinfo {title} {Accurate nuclear radii and binding energies from a chiral interaction},\ }\href {https://doi.org/10.1103/PhysRevC.91.051301} {\bibfield  {journal} {\bibinfo  {journal} {Phys. Rev. C}\ }\textbf {\bibinfo {volume} {91}},\ \bibinfo {pages} {051301} (\bibinfo {year} {2015})}\BibitemShut {NoStop}%
\bibitem [{\citenamefont {Piarulli}\ \emph {et~al.}(2016)\citenamefont {Piarulli}, \citenamefont {Girlanda}, \citenamefont {Schiavilla}, \citenamefont {Kievsky}, \citenamefont {Lovato}, \citenamefont {Marcucci}, \citenamefont {Pieper}, \citenamefont {Viviani},\ and\ \citenamefont {Wiringa}}]{piarulli2016}%
  \BibitemOpen
  \bibfield  {author} {\bibinfo {author} {\bibfnamefont {M.}~\bibnamefont {Piarulli}}, \bibinfo {author} {\bibfnamefont {L.}~\bibnamefont {Girlanda}}, \bibinfo {author} {\bibfnamefont {R.}~\bibnamefont {Schiavilla}}, \bibinfo {author} {\bibfnamefont {A.}~\bibnamefont {Kievsky}}, \bibinfo {author} {\bibfnamefont {A.}~\bibnamefont {Lovato}}, \bibinfo {author} {\bibfnamefont {L.~E.}\ \bibnamefont {Marcucci}}, \bibinfo {author} {\bibfnamefont {S.~C.}\ \bibnamefont {Pieper}}, \bibinfo {author} {\bibfnamefont {M.}~\bibnamefont {Viviani}},\ and\ \bibinfo {author} {\bibfnamefont {R.~B.}\ \bibnamefont {Wiringa}},\ }\bibfield  {title} {\bibinfo {title} {Local chiral potentials with $\mathrm{\ensuremath{\Delta}}$-intermediate states and the structure of light nuclei},\ }\href {https://doi.org/10.1103/PhysRevC.94.054007} {\bibfield  {journal} {\bibinfo  {journal} {Phys. Rev. C}\ }\textbf {\bibinfo {volume} {94}},\ \bibinfo {pages} {054007} (\bibinfo {year} {2016})}\BibitemShut {NoStop}%
\bibitem [{\citenamefont {Hammer}\ \emph {et~al.}(2020)\citenamefont {Hammer}, \citenamefont {K\"onig},\ and\ \citenamefont {van Kolck}}]{Hammer:2019poc}%
  \BibitemOpen
  \bibfield  {author} {\bibinfo {author} {\bibfnamefont {H.~W.}\ \bibnamefont {Hammer}}, \bibinfo {author} {\bibfnamefont {S.}~\bibnamefont {K\"onig}},\ and\ \bibinfo {author} {\bibfnamefont {U.}~\bibnamefont {van Kolck}},\ }\bibfield  {title} {\bibinfo {title} {{Nuclear effective field theory: Status and perspectives}},\ }\href {https://doi.org/10.1103/RevModPhys.92.025004} {\bibfield  {journal} {\bibinfo  {journal} {Rev. Mod. Phys.}\ }\textbf {\bibinfo {volume} {92}},\ \bibinfo {pages} {025004} (\bibinfo {year} {2020})}\BibitemShut {NoStop}%
\bibitem [{\citenamefont {Reinert}\ \emph {et~al.}(2018)\citenamefont {Reinert}, \citenamefont {Krebs},\ and\ \citenamefont {Epelbaum}}]{reinert2018}%
  \BibitemOpen
  \bibfield  {author} {\bibinfo {author} {\bibfnamefont {P.}~\bibnamefont {Reinert}}, \bibinfo {author} {\bibfnamefont {H.}~\bibnamefont {Krebs}},\ and\ \bibinfo {author} {\bibfnamefont {E.}~\bibnamefont {Epelbaum}},\ }\bibfield  {title} {\bibinfo {title} {Semilocal momentum-space regularized chiral two-nucleon potentials up to fifth order},\ }\href {https://doi.org/10.1140/epja/i2018-12516-4} {\bibfield  {journal} {\bibinfo  {journal} {Eur. Phys. J. A}\ }\textbf {\bibinfo {volume} {54}},\ \bibinfo {pages} {86} (\bibinfo {year} {2018})}\BibitemShut {NoStop}%
\bibitem [{\citenamefont {Maris}\ \emph {et~al.}(2021)\citenamefont {Maris}, \citenamefont {Epelbaum}, \citenamefont {Furnstahl}, \citenamefont {Golak}, \citenamefont {Hebeler}, \citenamefont {H\"uther}, \citenamefont {Kamada}, \citenamefont {Krebs}, \citenamefont {Mei\ss{}ner}, \citenamefont {Melendez}, \citenamefont {Nogga}, \citenamefont {Reinert}, \citenamefont {Roth}, \citenamefont {Skibi\ifmmode~\acute{n}\else \'{n}\fi{}ski}, \citenamefont {Soloviov}, \citenamefont {Topolnicki}, \citenamefont {Vary}, \citenamefont {Volkotrub}, \citenamefont {Wita\l{}a},\ and\ \citenamefont {Wolfgruber}}]{maris2021}%
  \BibitemOpen
  \bibfield  {author} {\bibinfo {author} {\bibfnamefont {P.}~\bibnamefont {Maris}}, \bibinfo {author} {\bibfnamefont {E.}~\bibnamefont {Epelbaum}}, \bibinfo {author} {\bibfnamefont {R.~J.}\ \bibnamefont {Furnstahl}}, \bibinfo {author} {\bibfnamefont {J.}~\bibnamefont {Golak}}, \bibinfo {author} {\bibfnamefont {K.}~\bibnamefont {Hebeler}}, \bibinfo {author} {\bibfnamefont {T.}~\bibnamefont {H\"uther}}, \bibinfo {author} {\bibfnamefont {H.}~\bibnamefont {Kamada}}, \bibinfo {author} {\bibfnamefont {H.}~\bibnamefont {Krebs}}, \bibinfo {author} {\bibfnamefont {U.-G.}\ \bibnamefont {Mei\ss{}ner}}, \bibinfo {author} {\bibfnamefont {J.~A.}\ \bibnamefont {Melendez}}, \bibinfo {author} {\bibfnamefont {A.}~\bibnamefont {Nogga}}, \bibinfo {author} {\bibfnamefont {P.}~\bibnamefont {Reinert}}, \bibinfo {author} {\bibfnamefont {R.}~\bibnamefont {Roth}}, \bibinfo {author} {\bibfnamefont {R.}~\bibnamefont {Skibi\ifmmode~\acute{n}\else \'{n}\fi{}ski}}, \bibinfo {author} {\bibfnamefont {V.}~\bibnamefont {Soloviov}}, \bibinfo
  {author} {\bibfnamefont {K.}~\bibnamefont {Topolnicki}}, \bibinfo {author} {\bibfnamefont {J.~P.}\ \bibnamefont {Vary}}, \bibinfo {author} {\bibfnamefont {Y.}~\bibnamefont {Volkotrub}}, \bibinfo {author} {\bibfnamefont {H.}~\bibnamefont {Wita\l{}a}},\ and\ \bibinfo {author} {\bibfnamefont {T.}~\bibnamefont {Wolfgruber}} (\bibinfo {collaboration} {LENPIC Collaboration}),\ }\bibfield  {title} {\bibinfo {title} {Light nuclei with semilocal momentum-space regularized chiral interactions up to third order},\ }\href {https://doi.org/10.1103/PhysRevC.103.054001} {\bibfield  {journal} {\bibinfo  {journal} {Phys. Rev. C}\ }\textbf {\bibinfo {volume} {103}},\ \bibinfo {pages} {054001} (\bibinfo {year} {2021})}\BibitemShut {NoStop}%
\bibitem [{\citenamefont {{Hagen}}\ \emph {et~al.}(2016)\citenamefont {{Hagen}}, \citenamefont {{Ekstr{\"o}m}}, \citenamefont {{Forss{\'e}n}}, \citenamefont {{Jansen}}, \citenamefont {{Nazarewicz}}, \citenamefont {{Papenbrock}}, \citenamefont {{Wendt}}, \citenamefont {{Bacca}}, \citenamefont {{Barnea}}, \citenamefont {{Carlsson}}, \citenamefont {{Drischler}}, \citenamefont {{Hebeler}}, \citenamefont {{Hjorth-Jensen}}, \citenamefont {{Miorelli}}, \citenamefont {{Orlandini}}, \citenamefont {{Schwenk}},\ and\ \citenamefont {{Simonis}}}]{hagen2015}%
  \BibitemOpen
  \bibfield  {author} {\bibinfo {author} {\bibfnamefont {G.}~\bibnamefont {{Hagen}}}, \bibinfo {author} {\bibfnamefont {A.}~\bibnamefont {{Ekstr{\"o}m}}}, \bibinfo {author} {\bibfnamefont {C.}~\bibnamefont {{Forss{\'e}n}}}, \bibinfo {author} {\bibfnamefont {G.~R.}\ \bibnamefont {{Jansen}}}, \bibinfo {author} {\bibfnamefont {W.}~\bibnamefont {{Nazarewicz}}}, \bibinfo {author} {\bibfnamefont {T.}~\bibnamefont {{Papenbrock}}}, \bibinfo {author} {\bibfnamefont {K.~A.}\ \bibnamefont {{Wendt}}}, \bibinfo {author} {\bibfnamefont {S.}~\bibnamefont {{Bacca}}}, \bibinfo {author} {\bibfnamefont {N.}~\bibnamefont {{Barnea}}}, \bibinfo {author} {\bibfnamefont {B.}~\bibnamefont {{Carlsson}}}, \bibinfo {author} {\bibfnamefont {C.}~\bibnamefont {{Drischler}}}, \bibinfo {author} {\bibfnamefont {K.}~\bibnamefont {{Hebeler}}}, \bibinfo {author} {\bibfnamefont {M.}~\bibnamefont {{Hjorth-Jensen}}}, \bibinfo {author} {\bibfnamefont {M.}~\bibnamefont {{Miorelli}}}, \bibinfo {author} {\bibfnamefont {G.}~\bibnamefont {{Orlandini}}},
  \bibinfo {author} {\bibfnamefont {A.}~\bibnamefont {{Schwenk}}},\ and\ \bibinfo {author} {\bibfnamefont {J.}~\bibnamefont {{Simonis}}},\ }\bibfield  {title} {\bibinfo {title} {{Neutron and weak-charge distributions of the $^{48}$Ca nucleus}},\ }\href {https://doi.org/10.1038/nphys3529} {\bibfield  {journal} {\bibinfo  {journal} {Nat. Phys.}\ }\textbf {\bibinfo {volume} {12}},\ \bibinfo {pages} {186} (\bibinfo {year} {2016})}\BibitemShut {NoStop}%
\bibitem [{\citenamefont {Hagen}\ \emph {et~al.}(2016)\citenamefont {Hagen}, \citenamefont {Jansen},\ and\ \citenamefont {Papenbrock}}]{hagen2017}%
  \BibitemOpen
  \bibfield  {author} {\bibinfo {author} {\bibfnamefont {G.}~\bibnamefont {Hagen}}, \bibinfo {author} {\bibfnamefont {G.~R.}\ \bibnamefont {Jansen}},\ and\ \bibinfo {author} {\bibfnamefont {T.}~\bibnamefont {Papenbrock}},\ }\bibfield  {title} {\bibinfo {title} {Structure of $^{78}\mathrm{Ni}$ from first-principles computations},\ }\href {https://doi.org/10.1103/PhysRevLett.117.172501} {\bibfield  {journal} {\bibinfo  {journal} {Phys. Rev. Lett.}\ }\textbf {\bibinfo {volume} {117}},\ \bibinfo {pages} {172501} (\bibinfo {year} {2016})}\BibitemShut {NoStop}%
\bibitem [{\citenamefont {Morris}\ \emph {et~al.}(2018)\citenamefont {Morris}, \citenamefont {Simonis}, \citenamefont {Stroberg}, \citenamefont {Stumpf}, \citenamefont {Hagen}, \citenamefont {Holt}, \citenamefont {Jansen}, \citenamefont {Papenbrock}, \citenamefont {Roth},\ and\ \citenamefont {Schwenk}}]{morris2018}%
  \BibitemOpen
  \bibfield  {author} {\bibinfo {author} {\bibfnamefont {T.~D.}\ \bibnamefont {Morris}}, \bibinfo {author} {\bibfnamefont {J.}~\bibnamefont {Simonis}}, \bibinfo {author} {\bibfnamefont {S.~R.}\ \bibnamefont {Stroberg}}, \bibinfo {author} {\bibfnamefont {C.}~\bibnamefont {Stumpf}}, \bibinfo {author} {\bibfnamefont {G.}~\bibnamefont {Hagen}}, \bibinfo {author} {\bibfnamefont {J.~D.}\ \bibnamefont {Holt}}, \bibinfo {author} {\bibfnamefont {G.~R.}\ \bibnamefont {Jansen}}, \bibinfo {author} {\bibfnamefont {T.}~\bibnamefont {Papenbrock}}, \bibinfo {author} {\bibfnamefont {R.}~\bibnamefont {Roth}},\ and\ \bibinfo {author} {\bibfnamefont {A.}~\bibnamefont {Schwenk}},\ }\bibfield  {title} {\bibinfo {title} {Structure of the lightest tin isotopes},\ }\href {https://doi.org/10.1103/PhysRevLett.120.152503} {\bibfield  {journal} {\bibinfo  {journal} {Phys. Rev. Lett.}\ }\textbf {\bibinfo {volume} {120}},\ \bibinfo {pages} {152503} (\bibinfo {year} {2018})}\BibitemShut {NoStop}%
\bibitem [{\citenamefont {Bonaiti}\ \emph {et~al.}()\citenamefont {Bonaiti}, \citenamefont {Hagen},\ and\ \citenamefont {Papenbrock}}]{Bonaiti:2025bsb}%
  \BibitemOpen
  \bibfield  {author} {\bibinfo {author} {\bibfnamefont {F.}~\bibnamefont {Bonaiti}}, \bibinfo {author} {\bibfnamefont {G.}~\bibnamefont {Hagen}},\ and\ \bibinfo {author} {\bibfnamefont {T.}~\bibnamefont {Papenbrock}},\ }\href@noop {} {\bibinfo {title} {{Structure of the doubly magic nuclei $^{208}$Pb and $^{266}$Pb from ab initio computations}}},\ \Eprint {https://arxiv.org/abs/2508.14217} {arXiv:2508.14217} \BibitemShut {NoStop}%
\bibitem [{\citenamefont {Gysbers}\ \emph {et~al.}(2019)\citenamefont {Gysbers}, \citenamefont {Hagen}, \citenamefont {Holt}, \citenamefont {Jansen}, \citenamefont {Morris}, \citenamefont {Navr{\'a}til}, \citenamefont {Papenbrock}, \citenamefont {Quaglioni}, \citenamefont {Schwenk}, \citenamefont {Stroberg},\ and\ \citenamefont {Wendt}}]{gysbers2019}%
  \BibitemOpen
  \bibfield  {author} {\bibinfo {author} {\bibfnamefont {P.}~\bibnamefont {Gysbers}}, \bibinfo {author} {\bibfnamefont {G.}~\bibnamefont {Hagen}}, \bibinfo {author} {\bibfnamefont {J.~D.}\ \bibnamefont {Holt}}, \bibinfo {author} {\bibfnamefont {G.~R.}\ \bibnamefont {Jansen}}, \bibinfo {author} {\bibfnamefont {T.~D.}\ \bibnamefont {Morris}}, \bibinfo {author} {\bibfnamefont {P.}~\bibnamefont {Navr{\'a}til}}, \bibinfo {author} {\bibfnamefont {T.}~\bibnamefont {Papenbrock}}, \bibinfo {author} {\bibfnamefont {S.}~\bibnamefont {Quaglioni}}, \bibinfo {author} {\bibfnamefont {A.}~\bibnamefont {Schwenk}}, \bibinfo {author} {\bibfnamefont {S.~R.}\ \bibnamefont {Stroberg}},\ and\ \bibinfo {author} {\bibfnamefont {K.~A.}\ \bibnamefont {Wendt}},\ }\bibfield  {title} {\bibinfo {title} {Discrepancy between experimental and theoretical $\beta$-decay rates resolved from first principles},\ }\href {https://doi.org/10.1038/s41567-019-0450-7} {\bibfield  {journal} {\bibinfo  {journal} {Nat. Phys.}\ }\textbf {\bibinfo {volume}
  {15}},\ \bibinfo {pages} {428} (\bibinfo {year} {2019})}\BibitemShut {NoStop}%
\bibitem [{\citenamefont {Hu}\ \emph {et~al.}(2022)\citenamefont {Hu}, \citenamefont {Jiang}, \citenamefont {Miyagi}, \citenamefont {Sun}, \citenamefont {Ekstr{\"o}m}, \citenamefont {Forss{\'e}n}, \citenamefont {Hagen}, \citenamefont {Holt}, \citenamefont {Papenbrock}, \citenamefont {Stroberg},\ and\ \citenamefont {Vernon}}]{hu2022}%
  \BibitemOpen
  \bibfield  {author} {\bibinfo {author} {\bibfnamefont {B.}~\bibnamefont {Hu}}, \bibinfo {author} {\bibfnamefont {W.}~\bibnamefont {Jiang}}, \bibinfo {author} {\bibfnamefont {T.}~\bibnamefont {Miyagi}}, \bibinfo {author} {\bibfnamefont {Z.}~\bibnamefont {Sun}}, \bibinfo {author} {\bibfnamefont {A.}~\bibnamefont {Ekstr{\"o}m}}, \bibinfo {author} {\bibfnamefont {C.}~\bibnamefont {Forss{\'e}n}}, \bibinfo {author} {\bibfnamefont {G.}~\bibnamefont {Hagen}}, \bibinfo {author} {\bibfnamefont {J.~D.}\ \bibnamefont {Holt}}, \bibinfo {author} {\bibfnamefont {T.}~\bibnamefont {Papenbrock}}, \bibinfo {author} {\bibfnamefont {S.~R.}\ \bibnamefont {Stroberg}},\ and\ \bibinfo {author} {\bibfnamefont {I.}~\bibnamefont {Vernon}},\ }\bibfield  {title} {\bibinfo {title} {Ab initio predictions link the neutron skin of $^{208}\mathrm{Pb}$ to nuclear forces},\ }\href {https://doi.org/10.1038/s41567-022-01715-8} {\bibfield  {journal} {\bibinfo  {journal} {Nat. Phys.}\ }\textbf {\bibinfo {volume} {18}},\ \bibinfo {pages} {1196}
  (\bibinfo {year} {2022})}\BibitemShut {NoStop}%
\bibitem [{\citenamefont {Sun}\ \emph {et~al.}(2025)\citenamefont {Sun}, \citenamefont {Ekstr\"om}, \citenamefont {Forss\'en}, \citenamefont {Hagen}, \citenamefont {Jansen},\ and\ \citenamefont {Papenbrock}}]{sun2025}%
  \BibitemOpen
  \bibfield  {author} {\bibinfo {author} {\bibfnamefont {Z.~H.}\ \bibnamefont {Sun}}, \bibinfo {author} {\bibfnamefont {A.}~\bibnamefont {Ekstr\"om}}, \bibinfo {author} {\bibfnamefont {C.}~\bibnamefont {Forss\'en}}, \bibinfo {author} {\bibfnamefont {G.}~\bibnamefont {Hagen}}, \bibinfo {author} {\bibfnamefont {G.~R.}\ \bibnamefont {Jansen}},\ and\ \bibinfo {author} {\bibfnamefont {T.}~\bibnamefont {Papenbrock}},\ }\bibfield  {title} {\bibinfo {title} {Multiscale physics of atomic nuclei from first principles},\ }\href {https://doi.org/10.1103/PhysRevX.15.011028} {\bibfield  {journal} {\bibinfo  {journal} {Phys. Rev. X}\ }\textbf {\bibinfo {volume} {15}},\ \bibinfo {pages} {011028} (\bibinfo {year} {2025})}\BibitemShut {NoStop}%
\bibitem [{\citenamefont {Ding}\ \emph {et~al.}(2026)\citenamefont {Ding}, \citenamefont {Wang}, \citenamefont {Yao}, \citenamefont {Hergert}, \citenamefont {Liang},\ and\ \citenamefont {Bogner}}]{ding2026}%
  \BibitemOpen
  \bibfield  {author} {\bibinfo {author} {\bibfnamefont {C.~R.}\ \bibnamefont {Ding}}, \bibinfo {author} {\bibfnamefont {C.~C.}\ \bibnamefont {Wang}}, \bibinfo {author} {\bibfnamefont {J.~M.}\ \bibnamefont {Yao}}, \bibinfo {author} {\bibfnamefont {H.}~\bibnamefont {Hergert}}, \bibinfo {author} {\bibfnamefont {H.~Z.}\ \bibnamefont {Liang}},\ and\ \bibinfo {author} {\bibfnamefont {S.~K.}\ \bibnamefont {Bogner}},\ }\bibfield  {title} {\bibinfo {title} {From spin to pseudospin symmetry: The origin of magic numbers in nuclear structure},\ }\href {https://doi.org/10.1103/8lzc-j1lx} {\bibfield  {journal} {\bibinfo  {journal} {Phys. Rev. Lett.}\ }\textbf {\bibinfo {volume} {136}},\ \bibinfo {pages} {052501} (\bibinfo {year} {2026})}\BibitemShut {NoStop}%
\bibitem [{\citenamefont {K{\"o}nig}\ \emph {et~al.}(2016)\citenamefont {K{\"o}nig}, \citenamefont {Grie{\ss}hammer}, \citenamefont {Hammer},\ and\ \citenamefont {van Kolck}}]{konig2016}%
  \BibitemOpen
  \bibfield  {author} {\bibinfo {author} {\bibfnamefont {S.}~\bibnamefont {K{\"o}nig}}, \bibinfo {author} {\bibfnamefont {H.~W.}\ \bibnamefont {Grie{\ss}hammer}}, \bibinfo {author} {\bibfnamefont {H.-W.}\ \bibnamefont {Hammer}},\ and\ \bibinfo {author} {\bibfnamefont {U.}~\bibnamefont {van Kolck}},\ }\bibfield  {title} {\bibinfo {title} {Effective theory of 3h and 3he},\ }\href {https://doi.org/10.1088/0954-3899/43/5/055106} {\bibfield  {journal} {\bibinfo  {journal} {J. Phys. G: Nucl. Part. Phys.}\ }\textbf {\bibinfo {volume} {43}},\ \bibinfo {pages} {055106} (\bibinfo {year} {2016})}\BibitemShut {NoStop}%
\bibitem [{\citenamefont {Elhatisari}\ \emph {et~al.}(2016)\citenamefont {Elhatisari}, \citenamefont {Li}, \citenamefont {Rokash}, \citenamefont {Alarc\'on}, \citenamefont {Du}, \citenamefont {Klein}, \citenamefont {Lu}, \citenamefont {Mei\ss{}ner}, \citenamefont {Epelbaum}, \citenamefont {Krebs}, \citenamefont {L\"ahde}, \citenamefont {Lee},\ and\ \citenamefont {Rupak}}]{elhatisari2016}%
  \BibitemOpen
  \bibfield  {author} {\bibinfo {author} {\bibfnamefont {S.}~\bibnamefont {Elhatisari}}, \bibinfo {author} {\bibfnamefont {N.}~\bibnamefont {Li}}, \bibinfo {author} {\bibfnamefont {A.}~\bibnamefont {Rokash}}, \bibinfo {author} {\bibfnamefont {J.~M.}\ \bibnamefont {Alarc\'on}}, \bibinfo {author} {\bibfnamefont {D.}~\bibnamefont {Du}}, \bibinfo {author} {\bibfnamefont {N.}~\bibnamefont {Klein}}, \bibinfo {author} {\bibfnamefont {B.-N.}\ \bibnamefont {Lu}}, \bibinfo {author} {\bibfnamefont {U.-G.}\ \bibnamefont {Mei\ss{}ner}}, \bibinfo {author} {\bibfnamefont {E.}~\bibnamefont {Epelbaum}}, \bibinfo {author} {\bibfnamefont {H.}~\bibnamefont {Krebs}}, \bibinfo {author} {\bibfnamefont {T.~A.}\ \bibnamefont {L\"ahde}}, \bibinfo {author} {\bibfnamefont {D.}~\bibnamefont {Lee}},\ and\ \bibinfo {author} {\bibfnamefont {G.}~\bibnamefont {Rupak}},\ }\bibfield  {title} {\bibinfo {title} {Nuclear binding near a quantum phase transition},\ }\href {https://doi.org/10.1103/PhysRevLett.117.132501} {\bibfield  {journal}
  {\bibinfo  {journal} {Phys. Rev. Lett.}\ }\textbf {\bibinfo {volume} {117}},\ \bibinfo {pages} {132501} (\bibinfo {year} {2016})}\BibitemShut {NoStop}%
\bibitem [{\citenamefont {Elhatisari}\ \emph {et~al.}(2017)\citenamefont {Elhatisari}, \citenamefont {Epelbaum}, \citenamefont {Krebs}, \citenamefont {L\"ahde}, \citenamefont {Lee}, \citenamefont {Li}, \citenamefont {Lu}, \citenamefont {Mei\ss{}ner},\ and\ \citenamefont {Rupak}}]{elhatisari2017}%
  \BibitemOpen
  \bibfield  {author} {\bibinfo {author} {\bibfnamefont {S.}~\bibnamefont {Elhatisari}}, \bibinfo {author} {\bibfnamefont {E.}~\bibnamefont {Epelbaum}}, \bibinfo {author} {\bibfnamefont {H.}~\bibnamefont {Krebs}}, \bibinfo {author} {\bibfnamefont {T.~A.}\ \bibnamefont {L\"ahde}}, \bibinfo {author} {\bibfnamefont {D.}~\bibnamefont {Lee}}, \bibinfo {author} {\bibfnamefont {N.}~\bibnamefont {Li}}, \bibinfo {author} {\bibfnamefont {B.-N.}\ \bibnamefont {Lu}}, \bibinfo {author} {\bibfnamefont {U.-G.}\ \bibnamefont {Mei\ss{}ner}},\ and\ \bibinfo {author} {\bibfnamefont {G.}~\bibnamefont {Rupak}},\ }\bibfield  {title} {\bibinfo {title} {Ab initio calculations of the isotopic dependence of nuclear clustering},\ }\href {https://doi.org/10.1103/PhysRevLett.119.222505} {\bibfield  {journal} {\bibinfo  {journal} {Phys. Rev. Lett.}\ }\textbf {\bibinfo {volume} {119}},\ \bibinfo {pages} {222505} (\bibinfo {year} {2017})}\BibitemShut {NoStop}%
\bibitem [{\citenamefont {Lu}\ \emph {et~al.}(2019)\citenamefont {Lu}, \citenamefont {Li}, \citenamefont {Elhatisari}, \citenamefont {Lee}, \citenamefont {Epelbaum},\ and\ \citenamefont {Mei{\ss}ner}}]{lu2019}%
  \BibitemOpen
  \bibfield  {author} {\bibinfo {author} {\bibfnamefont {B.-N.}\ \bibnamefont {Lu}}, \bibinfo {author} {\bibfnamefont {N.}~\bibnamefont {Li}}, \bibinfo {author} {\bibfnamefont {S.}~\bibnamefont {Elhatisari}}, \bibinfo {author} {\bibfnamefont {D.}~\bibnamefont {Lee}}, \bibinfo {author} {\bibfnamefont {E.}~\bibnamefont {Epelbaum}},\ and\ \bibinfo {author} {\bibfnamefont {U.-G.}\ \bibnamefont {Mei{\ss}ner}},\ }\bibfield  {title} {\bibinfo {title} {Essential elements for nuclear binding},\ }\href {https://doi.org/10.1016/j.physletb.2019.134863} {\bibfield  {journal} {\bibinfo  {journal} {Phys. Lett. B}\ }\textbf {\bibinfo {volume} {797}},\ \bibinfo {pages} {134863} (\bibinfo {year} {2019})}\BibitemShut {NoStop}%
\bibitem [{\citenamefont {Gnech}\ \emph {et~al.}(2024)\citenamefont {Gnech}, \citenamefont {Fore}, \citenamefont {Tropiano},\ and\ \citenamefont {Lovato}}]{gnech2024}%
  \BibitemOpen
  \bibfield  {author} {\bibinfo {author} {\bibfnamefont {A.}~\bibnamefont {Gnech}}, \bibinfo {author} {\bibfnamefont {B.}~\bibnamefont {Fore}}, \bibinfo {author} {\bibfnamefont {A.~J.}\ \bibnamefont {Tropiano}},\ and\ \bibinfo {author} {\bibfnamefont {A.}~\bibnamefont {Lovato}},\ }\bibfield  {title} {\bibinfo {title} {Distilling the essential elements of nuclear binding via neural-network quantum states},\ }\href {https://doi.org/10.1103/PhysRevLett.133.142501} {\bibfield  {journal} {\bibinfo  {journal} {Phys. Rev. Lett.}\ }\textbf {\bibinfo {volume} {133}},\ \bibinfo {pages} {142501} (\bibinfo {year} {2024})}\BibitemShut {NoStop}%
\bibitem [{\citenamefont {Drischler}\ \emph {et~al.}(2019)\citenamefont {Drischler}, \citenamefont {Hebeler},\ and\ \citenamefont {Schwenk}}]{drischler2019}%
  \BibitemOpen
  \bibfield  {author} {\bibinfo {author} {\bibfnamefont {C.}~\bibnamefont {Drischler}}, \bibinfo {author} {\bibfnamefont {K.}~\bibnamefont {Hebeler}},\ and\ \bibinfo {author} {\bibfnamefont {A.}~\bibnamefont {Schwenk}},\ }\bibfield  {title} {\bibinfo {title} {Chiral interactions up to next-to-next-to-next-to-leading order and nuclear saturation},\ }\href {https://doi.org/10.1103/PhysRevLett.122.042501} {\bibfield  {journal} {\bibinfo  {journal} {Phys. Rev. Lett.}\ }\textbf {\bibinfo {volume} {122}},\ \bibinfo {pages} {042501} (\bibinfo {year} {2019})}\BibitemShut {NoStop}%
\bibitem [{\citenamefont {Jiang}\ \emph {et~al.}(2020)\citenamefont {Jiang}, \citenamefont {Ekstr\"om}, \citenamefont {Forss\'en}, \citenamefont {Hagen}, \citenamefont {Jansen},\ and\ \citenamefont {Papenbrock}}]{jiang2020}%
  \BibitemOpen
  \bibfield  {author} {\bibinfo {author} {\bibfnamefont {W.~G.}\ \bibnamefont {Jiang}}, \bibinfo {author} {\bibfnamefont {A.}~\bibnamefont {Ekstr\"om}}, \bibinfo {author} {\bibfnamefont {C.}~\bibnamefont {Forss\'en}}, \bibinfo {author} {\bibfnamefont {G.}~\bibnamefont {Hagen}}, \bibinfo {author} {\bibfnamefont {G.~R.}\ \bibnamefont {Jansen}},\ and\ \bibinfo {author} {\bibfnamefont {T.}~\bibnamefont {Papenbrock}},\ }\bibfield  {title} {\bibinfo {title} {Accurate bulk properties of nuclei from {$A=2$} to $\ensuremath{\infty}$ from potentials with $\mathrm{\ensuremath{\Delta}}$ isobars},\ }\href {https://doi.org/10.1103/PhysRevC.102.054301} {\bibfield  {journal} {\bibinfo  {journal} {Phys. Rev. C}\ }\textbf {\bibinfo {volume} {102}},\ \bibinfo {pages} {054301} (\bibinfo {year} {2020})}\BibitemShut {NoStop}%
\bibitem [{\citenamefont {Arthuis}\ \emph {et~al.}(2024)\citenamefont {Arthuis}, \citenamefont {Hebeler},\ and\ \citenamefont {Schwenk}}]{arthuis2024}%
  \BibitemOpen
  \bibfield  {author} {\bibinfo {author} {\bibfnamefont {P.}~\bibnamefont {Arthuis}}, \bibinfo {author} {\bibfnamefont {K.}~\bibnamefont {Hebeler}},\ and\ \bibinfo {author} {\bibfnamefont {A.}~\bibnamefont {Schwenk}},\ }\href@noop {} {\bibinfo {title} {Neutron-rich nuclei and neutron skins from chiral low-resolution interactions}} (\bibinfo {year} {2024}),\ \bibinfo {note} {arXiv:2401.06675 [nucl-th]}\BibitemShut {NoStop}%
\bibitem [{\citenamefont {{Elhatisari}}\ \emph {et~al.}(2024)\citenamefont {{Elhatisari}}, \citenamefont {{Bovermann}}, \citenamefont {{Ma}}, \citenamefont {{Epelbaum}}, \citenamefont {{Frame}}, \citenamefont {{Hildenbrand}}, \citenamefont {{Kim}}, \citenamefont {{Kim}}, \citenamefont {{Krebs}}, \citenamefont {{L{\"a}hde}}, \citenamefont {{Lee}}, \citenamefont {{Li}}, \citenamefont {{Lu}}, \citenamefont {{Mei{\ss}ner}}, \citenamefont {{Rupak}}, \citenamefont {{Shen}}, \citenamefont {{Song}},\ and\ \citenamefont {{Stellin}}}]{elhatisari2024}%
  \BibitemOpen
  \bibfield  {author} {\bibinfo {author} {\bibfnamefont {S.}~\bibnamefont {{Elhatisari}}}, \bibinfo {author} {\bibfnamefont {L.}~\bibnamefont {{Bovermann}}}, \bibinfo {author} {\bibfnamefont {Y.-Z.}\ \bibnamefont {{Ma}}}, \bibinfo {author} {\bibfnamefont {E.}~\bibnamefont {{Epelbaum}}}, \bibinfo {author} {\bibfnamefont {D.}~\bibnamefont {{Frame}}}, \bibinfo {author} {\bibfnamefont {F.}~\bibnamefont {{Hildenbrand}}}, \bibinfo {author} {\bibfnamefont {M.}~\bibnamefont {{Kim}}}, \bibinfo {author} {\bibfnamefont {Y.}~\bibnamefont {{Kim}}}, \bibinfo {author} {\bibfnamefont {H.}~\bibnamefont {{Krebs}}}, \bibinfo {author} {\bibfnamefont {T.~A.}\ \bibnamefont {{L{\"a}hde}}}, \bibinfo {author} {\bibfnamefont {D.}~\bibnamefont {{Lee}}}, \bibinfo {author} {\bibfnamefont {N.}~\bibnamefont {{Li}}}, \bibinfo {author} {\bibfnamefont {B.-N.}\ \bibnamefont {{Lu}}}, \bibinfo {author} {\bibfnamefont {U.-G.}\ \bibnamefont {{Mei{\ss}ner}}}, \bibinfo {author} {\bibfnamefont {G.}~\bibnamefont {{Rupak}}}, \bibinfo {author}
  {\bibfnamefont {S.}~\bibnamefont {{Shen}}}, \bibinfo {author} {\bibfnamefont {Y.-H.}\ \bibnamefont {{Song}}},\ and\ \bibinfo {author} {\bibfnamefont {G.}~\bibnamefont {{Stellin}}},\ }\bibfield  {title} {\bibinfo {title} {{Wavefunction matching for solving quantum many-body problems}},\ }\href {https://doi.org/10.1038/s41586-024-07422-z} {\bibfield  {journal} {\bibinfo  {journal} {Nature}\ }\textbf {\bibinfo {volume} {630}},\ \bibinfo {pages} {59} (\bibinfo {year} {2024})}\BibitemShut {NoStop}%
\bibitem [{\citenamefont {{Bedaque}}\ and\ \citenamefont {{van Kolck}}(2002)}]{bedaque2002}%
  \BibitemOpen
  \bibfield  {author} {\bibinfo {author} {\bibfnamefont {P.~F.}\ \bibnamefont {{Bedaque}}}\ and\ \bibinfo {author} {\bibfnamefont {U.}~\bibnamefont {{van Kolck}}},\ }\bibfield  {title} {\bibinfo {title} {{Effective field theory for few-nucleon systems}},\ }\href {https://doi.org/10.1146/annurev.nucl.52.050102.090637} {\bibfield  {journal} {\bibinfo  {journal} {Annual Review of Nuclear and Particle Science}\ }\textbf {\bibinfo {volume} {52}},\ \bibinfo {pages} {339} (\bibinfo {year} {2002})},\ \Eprint {https://arxiv.org/abs/nucl-th/0203055} {nucl-th/0203055} \BibitemShut {NoStop}%
\bibitem [{\citenamefont {Hebeler}\ \emph {et~al.}(2011)\citenamefont {Hebeler}, \citenamefont {Bogner}, \citenamefont {Furnstahl}, \citenamefont {Nogga},\ and\ \citenamefont {Schwenk}}]{hebeler2011}%
  \BibitemOpen
  \bibfield  {author} {\bibinfo {author} {\bibfnamefont {K.}~\bibnamefont {Hebeler}}, \bibinfo {author} {\bibfnamefont {S.~K.}\ \bibnamefont {Bogner}}, \bibinfo {author} {\bibfnamefont {R.~J.}\ \bibnamefont {Furnstahl}}, \bibinfo {author} {\bibfnamefont {A.}~\bibnamefont {Nogga}},\ and\ \bibinfo {author} {\bibfnamefont {A.}~\bibnamefont {Schwenk}},\ }\bibfield  {title} {\bibinfo {title} {Improved nuclear matter calculations from chiral low-momentum interactions},\ }\href {https://doi.org/10.1103/PhysRevC.83.031301} {\bibfield  {journal} {\bibinfo  {journal} {Phys. Rev. C}\ }\textbf {\bibinfo {volume} {83}},\ \bibinfo {pages} {031301} (\bibinfo {year} {2011})}\BibitemShut {NoStop}%
\bibitem [{\citenamefont {Hagen}\ \emph {et~al.}(2007)\citenamefont {Hagen}, \citenamefont {Dean}, \citenamefont {Hjorth-Jensen}, \citenamefont {Papenbrock},\ and\ \citenamefont {Schwenk}}]{hagen2007b}%
  \BibitemOpen
  \bibfield  {author} {\bibinfo {author} {\bibfnamefont {G.}~\bibnamefont {Hagen}}, \bibinfo {author} {\bibfnamefont {D.~J.}\ \bibnamefont {Dean}}, \bibinfo {author} {\bibfnamefont {M.}~\bibnamefont {Hjorth-Jensen}}, \bibinfo {author} {\bibfnamefont {T.}~\bibnamefont {Papenbrock}},\ and\ \bibinfo {author} {\bibfnamefont {A.}~\bibnamefont {Schwenk}},\ }\bibfield  {title} {\bibinfo {title} {Benchmark calculations for $^{3}\mathrm{H}$, $^{4}\mathrm{He}$, $^{16}\mathrm{O}$, and $^{40}\mathrm{Ca}$ with ab initio coupled-cluster theory},\ }\href {https://doi.org/10.1103/PhysRevC.76.044305} {\bibfield  {journal} {\bibinfo  {journal} {Phys. Rev. C}\ }\textbf {\bibinfo {volume} {76}},\ \bibinfo {pages} {044305} (\bibinfo {year} {2007})}\BibitemShut {NoStop}%
\bibitem [{\citenamefont {Ekstr\"om}\ \emph {et~al.}(2013)\citenamefont {Ekstr\"om}, \citenamefont {Baardsen}, \citenamefont {Forss\'en}, \citenamefont {Hagen}, \citenamefont {Hjorth-Jensen}, \citenamefont {Jansen}, \citenamefont {Machleidt}, \citenamefont {Nazarewicz}, \citenamefont {Papenbrock}, \citenamefont {Sarich},\ and\ \citenamefont {Wild}}]{ekstrom2013}%
  \BibitemOpen
  \bibfield  {author} {\bibinfo {author} {\bibfnamefont {A.}~\bibnamefont {Ekstr\"om}}, \bibinfo {author} {\bibfnamefont {G.}~\bibnamefont {Baardsen}}, \bibinfo {author} {\bibfnamefont {C.}~\bibnamefont {Forss\'en}}, \bibinfo {author} {\bibfnamefont {G.}~\bibnamefont {Hagen}}, \bibinfo {author} {\bibfnamefont {M.}~\bibnamefont {Hjorth-Jensen}}, \bibinfo {author} {\bibfnamefont {G.~R.}\ \bibnamefont {Jansen}}, \bibinfo {author} {\bibfnamefont {R.}~\bibnamefont {Machleidt}}, \bibinfo {author} {\bibfnamefont {W.}~\bibnamefont {Nazarewicz}}, \bibinfo {author} {\bibfnamefont {T.}~\bibnamefont {Papenbrock}}, \bibinfo {author} {\bibfnamefont {J.}~\bibnamefont {Sarich}},\ and\ \bibinfo {author} {\bibfnamefont {S.~M.}\ \bibnamefont {Wild}},\ }\bibfield  {title} {\bibinfo {title} {Optimized chiral nucleon-nucleon interaction at next-to-next-to-leading order},\ }\href {https://doi.org/10.1103/PhysRevLett.110.192502} {\bibfield  {journal} {\bibinfo  {journal} {Phys. Rev. Lett.}\ }\textbf {\bibinfo {volume} {110}},\
  \bibinfo {pages} {192502} (\bibinfo {year} {2013})}\BibitemShut {NoStop}%
\bibitem [{\citenamefont {Carlson}\ \emph {et~al.}(1983)\citenamefont {Carlson}, \citenamefont {Pandharipande},\ and\ \citenamefont {Wiringa}}]{Carlson1983NPA_Urbana3NF}%
  \BibitemOpen
  \bibfield  {author} {\bibinfo {author} {\bibfnamefont {J.}~\bibnamefont {Carlson}}, \bibinfo {author} {\bibfnamefont {V.~R.}\ \bibnamefont {Pandharipande}},\ and\ \bibinfo {author} {\bibfnamefont {R.~B.}\ \bibnamefont {Wiringa}},\ }\bibfield  {title} {\bibinfo {title} {Three-nucleon interaction in 3-, 4- and {$\infty$}-body systems},\ }\href {https://doi.org/10.1016/0375-9474(83)90336-6} {\bibfield  {journal} {\bibinfo  {journal} {Nucl. Phys. A}\ }\textbf {\bibinfo {volume} {401}},\ \bibinfo {pages} {59} (\bibinfo {year} {1983})}\BibitemShut {NoStop}%
\bibitem [{\citenamefont {Pieper}\ \emph {et~al.}(2001)\citenamefont {Pieper}, \citenamefont {Pandharipande}, \citenamefont {Wiringa},\ and\ \citenamefont {Carlson}}]{pieper2001b}%
  \BibitemOpen
  \bibfield  {author} {\bibinfo {author} {\bibfnamefont {S.~C.}\ \bibnamefont {Pieper}}, \bibinfo {author} {\bibfnamefont {V.~R.}\ \bibnamefont {Pandharipande}}, \bibinfo {author} {\bibfnamefont {R.~B.}\ \bibnamefont {Wiringa}},\ and\ \bibinfo {author} {\bibfnamefont {J.}~\bibnamefont {Carlson}},\ }\bibfield  {title} {\bibinfo {title} {Realistic models of pion-exchange three-nucleon interactions},\ }\href {https://doi.org/10.1103/PhysRevC.64.014001} {\bibfield  {journal} {\bibinfo  {journal} {Phys. Rev. C}\ }\textbf {\bibinfo {volume} {64}},\ \bibinfo {pages} {014001} (\bibinfo {year} {2001})}\BibitemShut {NoStop}%
\bibitem [{\citenamefont {Hammer}\ \emph {et~al.}(2013)\citenamefont {Hammer}, \citenamefont {Nogga},\ and\ \citenamefont {Schwenk}}]{hammer2013}%
  \BibitemOpen
  \bibfield  {author} {\bibinfo {author} {\bibfnamefont {H.-W.}\ \bibnamefont {Hammer}}, \bibinfo {author} {\bibfnamefont {A.}~\bibnamefont {Nogga}},\ and\ \bibinfo {author} {\bibfnamefont {A.}~\bibnamefont {Schwenk}},\ }\bibfield  {title} {\bibinfo {title} {Colloquium: Three-body forces: From cold atoms to nuclei},\ }\href {https://doi.org/10.1103/RevModPhys.85.197} {\bibfield  {journal} {\bibinfo  {journal} {Rev. Mod. Phys.}\ }\textbf {\bibinfo {volume} {85}},\ \bibinfo {pages} {197} (\bibinfo {year} {2013})}\BibitemShut {NoStop}%
\bibitem [{\citenamefont {Hebeler}\ \emph {et~al.}(2015)\citenamefont {Hebeler}, \citenamefont {Holt}, \citenamefont {Men{\'e}ndez},\ and\ \citenamefont {Schwenk}}]{hebeler2015}%
  \BibitemOpen
  \bibfield  {author} {\bibinfo {author} {\bibfnamefont {K.}~\bibnamefont {Hebeler}}, \bibinfo {author} {\bibfnamefont {J.~D.}\ \bibnamefont {Holt}}, \bibinfo {author} {\bibfnamefont {J.}~\bibnamefont {Men{\'e}ndez}},\ and\ \bibinfo {author} {\bibfnamefont {A.}~\bibnamefont {Schwenk}},\ }\bibfield  {title} {\bibinfo {title} {Nuclear forces and their impact on neutron-rich nuclei and neutron-rich matter},\ }\href {https://doi.org/10.1146/annurev-nucl-102313-025446} {\bibfield  {journal} {\bibinfo  {journal} {Annu. Rev. Nucl. Part. Sci.}\ }\textbf {\bibinfo {volume} {65}},\ \bibinfo {pages} {457} (\bibinfo {year} {2015})}\BibitemShut {NoStop}%
\bibitem [{\citenamefont {Hebeler}(2021)}]{hebeler2021}%
  \BibitemOpen
  \bibfield  {author} {\bibinfo {author} {\bibfnamefont {K.}~\bibnamefont {Hebeler}},\ }\bibfield  {title} {\bibinfo {title} {Three-nucleon forces: Implementation and applications to atomic nuclei and dense matter},\ }\href {https://doi.org/10.1016/j.physrep.2020.08.009} {\bibfield  {journal} {\bibinfo  {journal} {Phys. Rep.}\ }\textbf {\bibinfo {volume} {890}},\ \bibinfo {pages} {1} (\bibinfo {year} {2021})}\BibitemShut {NoStop}%
\bibitem [{\citenamefont {Dickhoff}\ and\ \citenamefont {Barbieri}(2004)}]{dickhoff2004}%
  \BibitemOpen
  \bibfield  {author} {\bibinfo {author} {\bibfnamefont {W.}~\bibnamefont {Dickhoff}}\ and\ \bibinfo {author} {\bibfnamefont {C.}~\bibnamefont {Barbieri}},\ }\bibfield  {title} {\bibinfo {title} {Self-consistent {G}reen's function method for nuclei and nuclear matter},\ }\href {https://doi.org/10.1016/j.ppnp.2004.02.038} {\bibfield  {journal} {\bibinfo  {journal} {Prog. Part. Nucl. Phys.}\ }\textbf {\bibinfo {volume} {52}},\ \bibinfo {pages} {377 } (\bibinfo {year} {2004})}\BibitemShut {NoStop}%
\bibitem [{\citenamefont {Barrett}\ \emph {et~al.}(2013)\citenamefont {Barrett}, \citenamefont {Navr{\'a}til},\ and\ \citenamefont {Vary}}]{barrett2013}%
  \BibitemOpen
  \bibfield  {author} {\bibinfo {author} {\bibfnamefont {B.~R.}\ \bibnamefont {Barrett}}, \bibinfo {author} {\bibfnamefont {P.}~\bibnamefont {Navr{\'a}til}},\ and\ \bibinfo {author} {\bibfnamefont {J.~P.}\ \bibnamefont {Vary}},\ }\bibfield  {title} {\bibinfo {title} {Ab initio no core shell model},\ }\href {https://doi.org/10.1016/j.ppnp.2012.10.003} {\bibfield  {journal} {\bibinfo  {journal} {Prog. Part. Nucl. Phys.}\ }\textbf {\bibinfo {volume} {69}},\ \bibinfo {pages} {131 } (\bibinfo {year} {2013})}\BibitemShut {NoStop}%
\bibitem [{\citenamefont {Som\`a}\ \emph {et~al.}(2013)\citenamefont {Som\`a}, \citenamefont {Barbieri},\ and\ \citenamefont {Duguet}}]{soma2013}%
  \BibitemOpen
  \bibfield  {author} {\bibinfo {author} {\bibfnamefont {V.}~\bibnamefont {Som\`a}}, \bibinfo {author} {\bibfnamefont {C.}~\bibnamefont {Barbieri}},\ and\ \bibinfo {author} {\bibfnamefont {T.}~\bibnamefont {Duguet}},\ }\bibfield  {title} {\bibinfo {title} {Ab initio {G}orkov-{G}reen's function calculations of open-shell nuclei},\ }\href {https://doi.org/10.1103/PhysRevC.87.011303} {\bibfield  {journal} {\bibinfo  {journal} {Phys. Rev. C}\ }\textbf {\bibinfo {volume} {87}},\ \bibinfo {pages} {011303} (\bibinfo {year} {2013})}\BibitemShut {NoStop}%
\bibitem [{\citenamefont {Hagen}\ \emph {et~al.}(2014{\natexlab{a}})\citenamefont {Hagen}, \citenamefont {Papenbrock}, \citenamefont {Hjorth-Jensen},\ and\ \citenamefont {Dean}}]{hagen2014}%
  \BibitemOpen
  \bibfield  {author} {\bibinfo {author} {\bibfnamefont {G.}~\bibnamefont {Hagen}}, \bibinfo {author} {\bibfnamefont {T.}~\bibnamefont {Papenbrock}}, \bibinfo {author} {\bibfnamefont {M.}~\bibnamefont {Hjorth-Jensen}},\ and\ \bibinfo {author} {\bibfnamefont {D.~J.}\ \bibnamefont {Dean}},\ }\bibfield  {title} {\bibinfo {title} {Coupled-cluster computations of atomic nuclei},\ }\href {https://doi.org/10.1088/0034-4885/77/9/096302} {\bibfield  {journal} {\bibinfo  {journal} {Rep. Prog. Phys.}\ }\textbf {\bibinfo {volume} {77}},\ \bibinfo {pages} {096302} (\bibinfo {year} {2014}{\natexlab{a}})}\BibitemShut {NoStop}%
\bibitem [{\citenamefont {Hergert}\ \emph {et~al.}(2016)\citenamefont {Hergert}, \citenamefont {Bogner}, \citenamefont {Morris}, \citenamefont {Schwenk},\ and\ \citenamefont {Tsukiyama}}]{hergert2016}%
  \BibitemOpen
  \bibfield  {author} {\bibinfo {author} {\bibfnamefont {H.}~\bibnamefont {Hergert}}, \bibinfo {author} {\bibfnamefont {S.~K.}\ \bibnamefont {Bogner}}, \bibinfo {author} {\bibfnamefont {T.~D.}\ \bibnamefont {Morris}}, \bibinfo {author} {\bibfnamefont {A.}~\bibnamefont {Schwenk}},\ and\ \bibinfo {author} {\bibfnamefont {K.}~\bibnamefont {Tsukiyama}},\ }\bibfield  {title} {\bibinfo {title} {The in-medium similarity renormalization group: A novel ab initio method for nuclei},\ }\href {https://doi.org/10.1016/j.physrep.2015.12.007} {\bibfield  {journal} {\bibinfo  {journal} {Phys. Rep.}\ }\textbf {\bibinfo {volume} {621}},\ \bibinfo {pages} {165 } (\bibinfo {year} {2016})}\BibitemShut {NoStop}%
\bibitem [{\citenamefont {Launey}\ \emph {et~al.}(2016)\citenamefont {Launey}, \citenamefont {Dytrych},\ and\ \citenamefont {Draayer}}]{launey2016}%
  \BibitemOpen
  \bibfield  {author} {\bibinfo {author} {\bibfnamefont {K.~D.}\ \bibnamefont {Launey}}, \bibinfo {author} {\bibfnamefont {T.}~\bibnamefont {Dytrych}},\ and\ \bibinfo {author} {\bibfnamefont {J.~P.}\ \bibnamefont {Draayer}},\ }\bibfield  {title} {\bibinfo {title} {Symmetry-guided large-scale shell-model theory},\ }\href {https://doi.org/10.1016/j.ppnp.2016.02.001} {\bibfield  {journal} {\bibinfo  {journal} {Prog. Part. Nucl. Phys.}\ }\textbf {\bibinfo {volume} {89}},\ \bibinfo {pages} {101} (\bibinfo {year} {2016})}\BibitemShut {NoStop}%
\bibitem [{\citenamefont {Stroberg}\ \emph {et~al.}(2019)\citenamefont {Stroberg}, \citenamefont {Hergert}, \citenamefont {Bogner},\ and\ \citenamefont {Holt}}]{stroberg2019}%
  \BibitemOpen
  \bibfield  {author} {\bibinfo {author} {\bibfnamefont {S.~R.}\ \bibnamefont {Stroberg}}, \bibinfo {author} {\bibfnamefont {H.}~\bibnamefont {Hergert}}, \bibinfo {author} {\bibfnamefont {S.~K.}\ \bibnamefont {Bogner}},\ and\ \bibinfo {author} {\bibfnamefont {J.~D.}\ \bibnamefont {Holt}},\ }\bibfield  {title} {\bibinfo {title} {{Nonempirical Interactions for the Nuclear Shell Model: An Update}},\ }\href {https://doi.org/10.1146/annurev-nucl-101917-021120} {\bibfield  {journal} {\bibinfo  {journal} {Annu. Rev. Nucl. Part. Sci.}\ }\textbf {\bibinfo {volume} {69}},\ \bibinfo {pages} {307} (\bibinfo {year} {2019})}\BibitemShut {NoStop}%
\bibitem [{\citenamefont {Tichai}\ \emph {et~al.}(2020)\citenamefont {Tichai}, \citenamefont {Roth},\ and\ \citenamefont {Duguet}}]{tichai2020}%
  \BibitemOpen
  \bibfield  {author} {\bibinfo {author} {\bibfnamefont {A.}~\bibnamefont {Tichai}}, \bibinfo {author} {\bibfnamefont {R.}~\bibnamefont {Roth}},\ and\ \bibinfo {author} {\bibfnamefont {T.}~\bibnamefont {Duguet}},\ }\bibfield  {title} {\bibinfo {title} {Many-body perturbation theories for finite nuclei},\ }\href {https://doi.org/10.3389/fphy.2020.00164} {\bibfield  {journal} {\bibinfo  {journal} {Front. Phys.}\ }\textbf {\bibinfo {volume} {8}},\ \bibinfo {pages} {164} (\bibinfo {year} {2020})}\BibitemShut {NoStop}%
\bibitem [{\citenamefont {Heinz}\ \emph {et~al.}(2021)\citenamefont {Heinz}, \citenamefont {Tichai}, \citenamefont {Hoppe}, \citenamefont {Hebeler},\ and\ \citenamefont {Schwenk}}]{heinz2021}%
  \BibitemOpen
  \bibfield  {author} {\bibinfo {author} {\bibfnamefont {M.}~\bibnamefont {Heinz}}, \bibinfo {author} {\bibfnamefont {A.}~\bibnamefont {Tichai}}, \bibinfo {author} {\bibfnamefont {J.}~\bibnamefont {Hoppe}}, \bibinfo {author} {\bibfnamefont {K.}~\bibnamefont {Hebeler}},\ and\ \bibinfo {author} {\bibfnamefont {A.}~\bibnamefont {Schwenk}},\ }\bibfield  {title} {\bibinfo {title} {In-medium similarity renormalization group with three-body operators},\ }\href {https://doi.org/10.1103/PhysRevC.103.044318} {\bibfield  {journal} {\bibinfo  {journal} {Phys. Rev. C}\ }\textbf {\bibinfo {volume} {103}},\ \bibinfo {pages} {044318} (\bibinfo {year} {2021})}\BibitemShut {NoStop}%
\bibitem [{\citenamefont {{Piarulli}}\ \emph {et~al.}(2017)\citenamefont {{Piarulli}}, \citenamefont {{Baroni}}, \citenamefont {{Girlanda}}, \citenamefont {{Kievsky}}, \citenamefont {{Lovato}}, \citenamefont {{Lusk}}, \citenamefont {{Marcucci}}, \citenamefont {{Pieper}}, \citenamefont {{Schiavilla}}, \citenamefont {{Viviani}},\ and\ \citenamefont {{Wiringa}}}]{piarulli2017}%
  \BibitemOpen
  \bibfield  {author} {\bibinfo {author} {\bibfnamefont {M.}~\bibnamefont {{Piarulli}}}, \bibinfo {author} {\bibfnamefont {A.}~\bibnamefont {{Baroni}}}, \bibinfo {author} {\bibfnamefont {L.}~\bibnamefont {{Girlanda}}}, \bibinfo {author} {\bibfnamefont {A.}~\bibnamefont {{Kievsky}}}, \bibinfo {author} {\bibfnamefont {A.}~\bibnamefont {{Lovato}}}, \bibinfo {author} {\bibfnamefont {E.}~\bibnamefont {{Lusk}}}, \bibinfo {author} {\bibfnamefont {L.~E.}\ \bibnamefont {{Marcucci}}}, \bibinfo {author} {\bibfnamefont {S.~C.}\ \bibnamefont {{Pieper}}}, \bibinfo {author} {\bibfnamefont {R.}~\bibnamefont {{Schiavilla}}}, \bibinfo {author} {\bibfnamefont {M.}~\bibnamefont {{Viviani}}},\ and\ \bibinfo {author} {\bibfnamefont {R.~B.}\ \bibnamefont {{Wiringa}}},\ }\bibfield  {title} {\bibinfo {title} {{Light-nuclei spectra from chiral dynamics}},\ }\href {http://adsabs.harvard.edu/abs/2017arXiv170702883P} {\bibfield  {journal} {\bibinfo  {journal} {ArXiv e-prints}\ } (\bibinfo {year} {2017})},\ \Eprint
  {https://arxiv.org/abs/1707.02883} {arXiv:1707.02883 [nucl-th]} \BibitemShut {NoStop}%
\bibitem [{\citenamefont {Rothman}\ \emph {et~al.}(2026{\natexlab{a}})\citenamefont {Rothman}, \citenamefont {Johnson-Toth}, \citenamefont {Hagen}, \citenamefont {Heinz},\ and\ \citenamefont {Papenbrock}}]{Rothman:2025uza}%
  \BibitemOpen
  \bibfield  {author} {\bibinfo {author} {\bibfnamefont {M.}~\bibnamefont {Rothman}}, \bibinfo {author} {\bibfnamefont {B.}~\bibnamefont {Johnson-Toth}}, \bibinfo {author} {\bibfnamefont {G.}~\bibnamefont {Hagen}}, \bibinfo {author} {\bibfnamefont {M.}~\bibnamefont {Heinz}},\ and\ \bibinfo {author} {\bibfnamefont {T.}~\bibnamefont {Papenbrock}},\ }\bibfield  {title} {\bibinfo {title} {{NuLattice: Ab initio computations of atomic nuclei on lattices}},\ }\href {https://doi.org/10.1140/epja/s10050-025-01764-6} {\bibfield  {journal} {\bibinfo  {journal} {Eur. Phys. J. A}\ }\textbf {\bibinfo {volume} {62}},\ \bibinfo {pages} {28} (\bibinfo {year} {2026}{\natexlab{a}})}\BibitemShut {NoStop}%
\bibitem [{\citenamefont {Wang}\ \emph {et~al.}(2021)\citenamefont {Wang}, \citenamefont {Huang}, \citenamefont {Kondev}, \citenamefont {Audi},\ and\ \citenamefont {Naimi}}]{wang2021}%
  \BibitemOpen
  \bibfield  {author} {\bibinfo {author} {\bibfnamefont {M.}~\bibnamefont {Wang}}, \bibinfo {author} {\bibfnamefont {W.~J.}\ \bibnamefont {Huang}}, \bibinfo {author} {\bibfnamefont {F.~G.}\ \bibnamefont {Kondev}}, \bibinfo {author} {\bibfnamefont {G.}~\bibnamefont {Audi}},\ and\ \bibinfo {author} {\bibfnamefont {S.}~\bibnamefont {Naimi}},\ }\bibfield  {title} {\bibinfo {title} {{The AME 2020 atomic mass evaluation (II). Tables, graphs and references}},\ }\href {https://doi.org/10.1088/1674-1137/abddaf} {\bibfield  {journal} {\bibinfo  {journal} {Chin. Phys. C}\ }\textbf {\bibinfo {volume} {45}},\ \bibinfo {pages} {030003} (\bibinfo {year} {2021})}\BibitemShut {NoStop}%
\bibitem [{Note1()}]{Note1}%
  \BibitemOpen
  \bibinfo {note} {The auxiliary field Monte Carlo calculations of Refs.~\cite {epelbaum2010,epelbaum2010b,epelbaum2011,epelbaum2012,epelbaum2013,epelbaum2014,elhatisari2016,elhatisari2017} solved the many-body problem by applying the transfer matrix formalism~\cite {lee2009} with a discrete temporal lattice spacing $a_t=1/(150\protect \,\protect \mathrm {MeV})$. Such calculations are not equivalent to calculations with $a_t\to 0$ (see Supplemental Material). Based on our comparison to Refs.~\cite {elhatisari2016,elhatisari2017}, the differences are significant.}\BibitemShut {Stop}%
\bibitem [{\citenamefont {Gandolfi}\ \emph {et~al.}(2009)\citenamefont {Gandolfi}, \citenamefont {Illarionov}, \citenamefont {Schmidt}, \citenamefont {Pederiva},\ and\ \citenamefont {Fantoni}}]{gandolfi2009}%
  \BibitemOpen
  \bibfield  {author} {\bibinfo {author} {\bibfnamefont {S.}~\bibnamefont {Gandolfi}}, \bibinfo {author} {\bibfnamefont {A.~Y.}\ \bibnamefont {Illarionov}}, \bibinfo {author} {\bibfnamefont {K.~E.}\ \bibnamefont {Schmidt}}, \bibinfo {author} {\bibfnamefont {F.}~\bibnamefont {Pederiva}},\ and\ \bibinfo {author} {\bibfnamefont {S.}~\bibnamefont {Fantoni}},\ }\bibfield  {title} {\bibinfo {title} {Quantum {Monte Carlo} calculation of the equation of state of neutron matter},\ }\href {https://doi.org/10.1103/PhysRevC.79.054005} {\bibfield  {journal} {\bibinfo  {journal} {Phys. Rev. C}\ }\textbf {\bibinfo {volume} {79}},\ \bibinfo {pages} {054005} (\bibinfo {year} {2009})}\BibitemShut {NoStop}%
\bibitem [{\citenamefont {Hagen}\ \emph {et~al.}(2014{\natexlab{b}})\citenamefont {Hagen}, \citenamefont {Papenbrock}, \citenamefont {Ekstr\"om}, \citenamefont {Wendt}, \citenamefont {Baardsen}, \citenamefont {Gandolfi}, \citenamefont {Hjorth-Jensen},\ and\ \citenamefont {Horowitz}}]{hagen2013b}%
  \BibitemOpen
  \bibfield  {author} {\bibinfo {author} {\bibfnamefont {G.}~\bibnamefont {Hagen}}, \bibinfo {author} {\bibfnamefont {T.}~\bibnamefont {Papenbrock}}, \bibinfo {author} {\bibfnamefont {A.}~\bibnamefont {Ekstr\"om}}, \bibinfo {author} {\bibfnamefont {K.~A.}\ \bibnamefont {Wendt}}, \bibinfo {author} {\bibfnamefont {G.}~\bibnamefont {Baardsen}}, \bibinfo {author} {\bibfnamefont {S.}~\bibnamefont {Gandolfi}}, \bibinfo {author} {\bibfnamefont {M.}~\bibnamefont {Hjorth-Jensen}},\ and\ \bibinfo {author} {\bibfnamefont {C.~J.}\ \bibnamefont {Horowitz}},\ }\bibfield  {title} {\bibinfo {title} {Coupled-cluster calculations of nucleonic matter},\ }\href {https://doi.org/10.1103/PhysRevC.89.014319} {\bibfield  {journal} {\bibinfo  {journal} {Phys. Rev. C}\ }\textbf {\bibinfo {volume} {89}},\ \bibinfo {pages} {014319} (\bibinfo {year} {2014}{\natexlab{b}})}\BibitemShut {NoStop}%
\bibitem [{\citenamefont {Marino}\ \emph {et~al.}(2024)\citenamefont {Marino}, \citenamefont {Jiang},\ and\ \citenamefont {Novario}}]{marino2024}%
  \BibitemOpen
  \bibfield  {author} {\bibinfo {author} {\bibfnamefont {F.}~\bibnamefont {Marino}}, \bibinfo {author} {\bibfnamefont {W.~G.}\ \bibnamefont {Jiang}},\ and\ \bibinfo {author} {\bibfnamefont {S.~J.}\ \bibnamefont {Novario}},\ }\bibfield  {title} {\bibinfo {title} {Diagrammatic ab initio methods for infinite nuclear matter with modern chiral interactions},\ }\href {https://doi.org/10.1103/PhysRevC.110.054322} {\bibfield  {journal} {\bibinfo  {journal} {Phys. Rev. C}\ }\textbf {\bibinfo {volume} {110}},\ \bibinfo {pages} {054322} (\bibinfo {year} {2024})}\BibitemShut {NoStop}%
\bibitem [{\citenamefont {{Christina Colizza}}(2025)}]{wirecutter}%
  \BibitemOpen
  \bibfield  {author} {\bibinfo {author} {\bibnamefont {{Christina Colizza}}},\ }\href@noop {} {\bibinfo {title} {{The Anatomy of a Wirecutter Guide}}},\ \bibinfo {howpublished} {\url{https://www.nytimes.com/wirecutter/reviews/anatomy-of-a-guide/}} (\bibinfo {year} {2025}),\ \bibinfo {note} {accessed: 2026-04-09}\BibitemShut {NoStop}%
\bibitem [{\citenamefont {Elhatisari}(2026)}]{elhatisari2026}%
  \BibitemOpen
  \bibfield  {author} {\bibinfo {author} {\bibfnamefont {S.}~\bibnamefont {Elhatisari}},\ }\href@noop {} {\bibinfo {title} {{private communications}}} (\bibinfo {year} {2026})\BibitemShut {NoStop}%
\bibitem [{\citenamefont {Rothman}\ \emph {et~al.}(2026{\natexlab{b}})\citenamefont {Rothman}, \citenamefont {Hagen}, \citenamefont {Heinz},\ and\ \citenamefont {Papenbrock}}]{Rothman2026NuLatticev1.2}%
  \BibitemOpen
  \bibfield  {author} {\bibinfo {author} {\bibfnamefont {M.}~\bibnamefont {Rothman}}, \bibinfo {author} {\bibfnamefont {G.}~\bibnamefont {Hagen}}, \bibinfo {author} {\bibfnamefont {M.}~\bibnamefont {Heinz}},\ and\ \bibinfo {author} {\bibfnamefont {T.}~\bibnamefont {Papenbrock}},\ }\href {https://doi.org/10.5281/zenodo.20617783} {\bibinfo {title} {Nulattice v1.2: Saturation of nuclear binding from lattice hamiltonians}} (\bibinfo {year} {2026}{\natexlab{b}})\BibitemShut {NoStop}%
\bibitem [{\citenamefont {Wang}\ \emph {et~al.}()\citenamefont {Wang}, \citenamefont {Shi},\ and\ \citenamefont {Lu}}]{wang2026}%
  \BibitemOpen
  \bibfield  {author} {\bibinfo {author} {\bibfnamefont {C.-C.}\ \bibnamefont {Wang}}, \bibinfo {author} {\bibfnamefont {J.-A.}\ \bibnamefont {Shi}},\ and\ \bibinfo {author} {\bibfnamefont {B.-N.}\ \bibnamefont {Lu}},\ }\href@noop {} {\bibinfo {title} {{Cutoff-independent predictions from nuclear lattice effective field theory}}},\ \Eprint {https://arxiv.org/abs/2604.20681} {arXiv:2604.20681} \BibitemShut {NoStop}%
\bibitem [{\citenamefont {Agar}\ \emph {et~al.}()\citenamefont {Agar}, \citenamefont {Ren},\ and\ \citenamefont {Elhatisari}}]{agar2026}%
  \BibitemOpen
  \bibfield  {author} {\bibinfo {author} {\bibfnamefont {O.}~\bibnamefont {Agar}}, \bibinfo {author} {\bibfnamefont {Z.}~\bibnamefont {Ren}},\ and\ \bibinfo {author} {\bibfnamefont {S.}~\bibnamefont {Elhatisari}},\ }\href@noop {} {\bibinfo {title} {{From binding and saturation to criticality in nuclear matter with lattice effective field theory}}},\ \Eprint {https://arxiv.org/abs/2604.09154} {arXiv:2604.09154} \BibitemShut {NoStop}%
\bibitem [{\citenamefont {Nogga}\ \emph {et~al.}(2004)\citenamefont {Nogga}, \citenamefont {Bogner},\ and\ \citenamefont {Schwenk}}]{nogga2004}%
  \BibitemOpen
  \bibfield  {author} {\bibinfo {author} {\bibfnamefont {A.}~\bibnamefont {Nogga}}, \bibinfo {author} {\bibfnamefont {S.~K.}\ \bibnamefont {Bogner}},\ and\ \bibinfo {author} {\bibfnamefont {A.}~\bibnamefont {Schwenk}},\ }\bibfield  {title} {\bibinfo {title} {Low-momentum interaction in few-nucleon systems},\ }\href {https://doi.org/10.1103/PhysRevC.70.061002} {\bibfield  {journal} {\bibinfo  {journal} {Phys. Rev. C}\ }\textbf {\bibinfo {volume} {70}},\ \bibinfo {pages} {061002} (\bibinfo {year} {2004})}\BibitemShut {NoStop}%
\bibitem [{\citenamefont {{Bogner}}\ \emph {et~al.}(2005)\citenamefont {{Bogner}}, \citenamefont {{Schwenk}}, \citenamefont {{Furnstahl}},\ and\ \citenamefont {{Nogga}}}]{bogner2005}%
  \BibitemOpen
  \bibfield  {author} {\bibinfo {author} {\bibfnamefont {S.~K.}\ \bibnamefont {{Bogner}}}, \bibinfo {author} {\bibfnamefont {A.}~\bibnamefont {{Schwenk}}}, \bibinfo {author} {\bibfnamefont {R.~J.}\ \bibnamefont {{Furnstahl}}},\ and\ \bibinfo {author} {\bibfnamefont {A.}~\bibnamefont {{Nogga}}},\ }\bibfield  {title} {\bibinfo {title} {{Is nuclear matter perturbative with low-momentum interactions?}},\ }\href {https://doi.org/10.1016/j.nuclphysa.2005.08.024} {\bibfield  {journal} {\bibinfo  {journal} {Nucl. Phys. A}\ }\textbf {\bibinfo {volume} {763}},\ \bibinfo {pages} {59} (\bibinfo {year} {2005})},\ \Eprint {https://arxiv.org/abs/nucl-th/0504043} {nucl-th/0504043} \BibitemShut {NoStop}%
\bibitem [{\citenamefont {Bogner}\ \emph {et~al.}(2010)\citenamefont {Bogner}, \citenamefont {Furnstahl},\ and\ \citenamefont {Schwenk}}]{bogner2010}%
  \BibitemOpen
  \bibfield  {author} {\bibinfo {author} {\bibfnamefont {S.}~\bibnamefont {Bogner}}, \bibinfo {author} {\bibfnamefont {R.}~\bibnamefont {Furnstahl}},\ and\ \bibinfo {author} {\bibfnamefont {A.}~\bibnamefont {Schwenk}},\ }\bibfield  {title} {\bibinfo {title} {From low-momentum interactions to nuclear structure},\ }\href {https://doi.org/10.1016/j.ppnp.2010.03.001} {\bibfield  {journal} {\bibinfo  {journal} {Prog. Part. Nucl. Phys.}\ }\textbf {\bibinfo {volume} {65}},\ \bibinfo {pages} {94 } (\bibinfo {year} {2010})}\BibitemShut {NoStop}%
\bibitem [{\citenamefont {Alp}\ \emph {et~al.}(2025)\citenamefont {Alp}, \citenamefont {Dietz}, \citenamefont {Hebeler},\ and\ \citenamefont {Schwenk}}]{alp2025}%
  \BibitemOpen
  \bibfield  {author} {\bibinfo {author} {\bibfnamefont {F.}~\bibnamefont {Alp}}, \bibinfo {author} {\bibfnamefont {Y.}~\bibnamefont {Dietz}}, \bibinfo {author} {\bibfnamefont {K.}~\bibnamefont {Hebeler}},\ and\ \bibinfo {author} {\bibfnamefont {A.}~\bibnamefont {Schwenk}},\ }\bibfield  {title} {\bibinfo {title} {Equation of state and {Fermi} liquid properties of dense matter based on chiral effective field theory interactions},\ }\href {https://doi.org/10.1103/ls3l-dn1y} {\bibfield  {journal} {\bibinfo  {journal} {Phys. Rev. C}\ }\textbf {\bibinfo {volume} {112}},\ \bibinfo {pages} {055802} (\bibinfo {year} {2025})}\BibitemShut {NoStop}%
\bibitem [{\citenamefont {Niu}\ and\ \citenamefont {Lu}(2025)}]{niu2025}%
  \BibitemOpen
  \bibfield  {author} {\bibinfo {author} {\bibfnamefont {Z.-W.}\ \bibnamefont {Niu}}\ and\ \bibinfo {author} {\bibfnamefont {B.-N.}\ \bibnamefont {Lu}},\ }\bibfield  {title} {\bibinfo {title} {Sign-problem-free nuclear quantum {Monte Carlo} simulation},\ }\href {https://doi.org/10.1103/pn99-6dxt} {\bibfield  {journal} {\bibinfo  {journal} {Phys. Rev. Lett.}\ }\textbf {\bibinfo {volume} {135}},\ \bibinfo {pages} {222504} (\bibinfo {year} {2025})}\BibitemShut {NoStop}%
\bibitem [{\citenamefont {Lieb}(1976)}]{lieb1976}%
  \BibitemOpen
  \bibfield  {author} {\bibinfo {author} {\bibfnamefont {E.~H.}\ \bibnamefont {Lieb}},\ }\bibfield  {title} {\bibinfo {title} {The stability of matter},\ }\href {https://doi.org/10.1103/RevModPhys.48.553} {\bibfield  {journal} {\bibinfo  {journal} {Rev. Mod. Phys.}\ }\textbf {\bibinfo {volume} {48}},\ \bibinfo {pages} {553} (\bibinfo {year} {1976})}\BibitemShut {NoStop}%
\bibitem [{\citenamefont {Freer}\ \emph {et~al.}(2018)\citenamefont {Freer}, \citenamefont {Horiuchi}, \citenamefont {Kanada-En'yo}, \citenamefont {Lee},\ and\ \citenamefont {Mei\ss{}ner}}]{freer2018}%
  \BibitemOpen
  \bibfield  {author} {\bibinfo {author} {\bibfnamefont {M.}~\bibnamefont {Freer}}, \bibinfo {author} {\bibfnamefont {H.}~\bibnamefont {Horiuchi}}, \bibinfo {author} {\bibfnamefont {Y.}~\bibnamefont {Kanada-En'yo}}, \bibinfo {author} {\bibfnamefont {D.}~\bibnamefont {Lee}},\ and\ \bibinfo {author} {\bibfnamefont {U.-G.}\ \bibnamefont {Mei\ss{}ner}},\ }\bibfield  {title} {\bibinfo {title} {Microscopic clustering in light nuclei},\ }\href {https://doi.org/10.1103/RevModPhys.90.035004} {\bibfield  {journal} {\bibinfo  {journal} {Rev. Mod. Phys.}\ }\textbf {\bibinfo {volume} {90}},\ \bibinfo {pages} {035004} (\bibinfo {year} {2018})}\BibitemShut {NoStop}%
\bibitem [{\citenamefont {Otsuka}\ \emph {et~al.}(2026)\citenamefont {Otsuka}, \citenamefont {Volya},\ and\ \citenamefont {Itagaki}}]{Otsuka:2026vwf}%
  \BibitemOpen
  \bibfield  {author} {\bibinfo {author} {\bibfnamefont {T.}~\bibnamefont {Otsuka}}, \bibinfo {author} {\bibfnamefont {A.}~\bibnamefont {Volya}},\ and\ \bibinfo {author} {\bibfnamefont {N.}~\bibnamefont {Itagaki}},\ }\bibfield  {title} {\bibinfo {title} {{Theoretical studies of $\alpha $ clustering in nuclei and beyond}},\ }\href {https://doi.org/10.1140/epja/s10050-026-01856-x} {\bibfield  {journal} {\bibinfo  {journal} {Eur. Phys. J. A}\ }\textbf {\bibinfo {volume} {62}},\ \bibinfo {pages} {108} (\bibinfo {year} {2026})},\ \Eprint {https://arxiv.org/abs/2602.24175} {arXiv:2602.24175 [nucl-th]} \BibitemShut {NoStop}%
\bibitem [{\citenamefont {Epelbaum}\ \emph {et~al.}(2010{\natexlab{a}})\citenamefont {Epelbaum}, \citenamefont {Krebs}, \citenamefont {Lee},\ and\ \citenamefont {Mei{\ss}ner}}]{epelbaum2010}%
  \BibitemOpen
  \bibfield  {author} {\bibinfo {author} {\bibfnamefont {E.}~\bibnamefont {Epelbaum}}, \bibinfo {author} {\bibfnamefont {H.}~\bibnamefont {Krebs}}, \bibinfo {author} {\bibfnamefont {D.}~\bibnamefont {Lee}},\ and\ \bibinfo {author} {\bibfnamefont {U.-G.}\ \bibnamefont {Mei{\ss}ner}},\ }\bibfield  {title} {\bibinfo {title} {Lattice calculations for {A = 3 , 4, 6, 12} nuclei using chiral effective field theory},\ }\href {https://doi.org/10.1140/epja/i2010-11009-x} {\bibfield  {journal} {\bibinfo  {journal} {Eur. Phys. J. A}\ }\textbf {\bibinfo {volume} {45}},\ \bibinfo {pages} {335} (\bibinfo {year} {2010}{\natexlab{a}})}\BibitemShut {NoStop}%
\bibitem [{\citenamefont {Epelbaum}\ \emph {et~al.}(2010{\natexlab{b}})\citenamefont {Epelbaum}, \citenamefont {Krebs}, \citenamefont {Lee},\ and\ \citenamefont {Mei\ss{}ner}}]{epelbaum2010b}%
  \BibitemOpen
  \bibfield  {author} {\bibinfo {author} {\bibfnamefont {E.}~\bibnamefont {Epelbaum}}, \bibinfo {author} {\bibfnamefont {H.}~\bibnamefont {Krebs}}, \bibinfo {author} {\bibfnamefont {D.}~\bibnamefont {Lee}},\ and\ \bibinfo {author} {\bibfnamefont {U.-G.}\ \bibnamefont {Mei\ss{}ner}},\ }\bibfield  {title} {\bibinfo {title} {Lattice effective field theory calculations for {$A=3$, 4, 6, 12} nuclei},\ }\href {https://doi.org/10.1103/PhysRevLett.104.142501} {\bibfield  {journal} {\bibinfo  {journal} {Phys. Rev. Lett.}\ }\textbf {\bibinfo {volume} {104}},\ \bibinfo {pages} {142501} (\bibinfo {year} {2010}{\natexlab{b}})}\BibitemShut {NoStop}%
\bibitem [{\citenamefont {Epelbaum}\ \emph {et~al.}(2011)\citenamefont {Epelbaum}, \citenamefont {Krebs}, \citenamefont {Lee},\ and\ \citenamefont {Mei\ss{}ner}}]{epelbaum2011}%
  \BibitemOpen
  \bibfield  {author} {\bibinfo {author} {\bibfnamefont {E.}~\bibnamefont {Epelbaum}}, \bibinfo {author} {\bibfnamefont {H.}~\bibnamefont {Krebs}}, \bibinfo {author} {\bibfnamefont {D.}~\bibnamefont {Lee}},\ and\ \bibinfo {author} {\bibfnamefont {U.-G.}\ \bibnamefont {Mei\ss{}ner}},\ }\bibfield  {title} {\bibinfo {title} {{\textit{Ab~Initio} Calculation of the {Hoyle} State}},\ }\href {https://doi.org/10.1103/PhysRevLett.106.192501} {\bibfield  {journal} {\bibinfo  {journal} {Phys. Rev. Lett.}\ }\textbf {\bibinfo {volume} {106}},\ \bibinfo {pages} {192501} (\bibinfo {year} {2011})}\BibitemShut {NoStop}%
\bibitem [{\citenamefont {Epelbaum}\ \emph {et~al.}(2012)\citenamefont {Epelbaum}, \citenamefont {Krebs}, \citenamefont {L\"ahde}, \citenamefont {Lee},\ and\ \citenamefont {Mei\ss{}ner}}]{epelbaum2012}%
  \BibitemOpen
  \bibfield  {author} {\bibinfo {author} {\bibfnamefont {E.}~\bibnamefont {Epelbaum}}, \bibinfo {author} {\bibfnamefont {H.}~\bibnamefont {Krebs}}, \bibinfo {author} {\bibfnamefont {T.~A.}\ \bibnamefont {L\"ahde}}, \bibinfo {author} {\bibfnamefont {D.}~\bibnamefont {Lee}},\ and\ \bibinfo {author} {\bibfnamefont {U.-G.}\ \bibnamefont {Mei\ss{}ner}},\ }\bibfield  {title} {\bibinfo {title} {Structure and rotations of the {Hoyle} state},\ }\href {https://doi.org/10.1103/PhysRevLett.109.252501} {\bibfield  {journal} {\bibinfo  {journal} {Phys. Rev. Lett.}\ }\textbf {\bibinfo {volume} {109}},\ \bibinfo {pages} {252501} (\bibinfo {year} {2012})}\BibitemShut {NoStop}%
\bibitem [{\citenamefont {Epelbaum}\ \emph {et~al.}(2013)\citenamefont {Epelbaum}, \citenamefont {Krebs}, \citenamefont {L\"ahde}, \citenamefont {Lee},\ and\ \citenamefont {Mei\ss{}ner}}]{epelbaum2013}%
  \BibitemOpen
  \bibfield  {author} {\bibinfo {author} {\bibfnamefont {E.}~\bibnamefont {Epelbaum}}, \bibinfo {author} {\bibfnamefont {H.}~\bibnamefont {Krebs}}, \bibinfo {author} {\bibfnamefont {T.~A.}\ \bibnamefont {L\"ahde}}, \bibinfo {author} {\bibfnamefont {D.}~\bibnamefont {Lee}},\ and\ \bibinfo {author} {\bibfnamefont {U.-G.}\ \bibnamefont {Mei\ss{}ner}},\ }\bibfield  {title} {\bibinfo {title} {Viability of carbon-based life as a function of the light quark mass},\ }\href {https://doi.org/10.1103/PhysRevLett.110.112502} {\bibfield  {journal} {\bibinfo  {journal} {Phys. Rev. Lett.}\ }\textbf {\bibinfo {volume} {110}},\ \bibinfo {pages} {112502} (\bibinfo {year} {2013})}\BibitemShut {NoStop}%
\bibitem [{\citenamefont {Epelbaum}\ \emph {et~al.}(2014)\citenamefont {Epelbaum}, \citenamefont {Krebs}, \citenamefont {L\"ahde}, \citenamefont {Lee}, \citenamefont {Mei\ss{}ner},\ and\ \citenamefont {Rupak}}]{epelbaum2014}%
  \BibitemOpen
  \bibfield  {author} {\bibinfo {author} {\bibfnamefont {E.}~\bibnamefont {Epelbaum}}, \bibinfo {author} {\bibfnamefont {H.}~\bibnamefont {Krebs}}, \bibinfo {author} {\bibfnamefont {T.~A.}\ \bibnamefont {L\"ahde}}, \bibinfo {author} {\bibfnamefont {D.}~\bibnamefont {Lee}}, \bibinfo {author} {\bibfnamefont {U.-G.}\ \bibnamefont {Mei\ss{}ner}},\ and\ \bibinfo {author} {\bibfnamefont {G.}~\bibnamefont {Rupak}},\ }\bibfield  {title} {\bibinfo {title} {Ab initio calculation of the spectrum and structure of {$^{16}$O}},\ }\href {https://doi.org/10.1103/PhysRevLett.112.102501} {\bibfield  {journal} {\bibinfo  {journal} {Phys. Rev. Lett.}\ }\textbf {\bibinfo {volume} {112}},\ \bibinfo {pages} {102501} (\bibinfo {year} {2014})}\BibitemShut {NoStop}%
\bibitem [{\citenamefont {Lee}(2009)}]{lee2009}%
  \BibitemOpen
  \bibfield  {author} {\bibinfo {author} {\bibfnamefont {D.}~\bibnamefont {Lee}},\ }\bibfield  {title} {\bibinfo {title} {Lattice simulations for few- and many-body systems},\ }\href {https://doi.org/10.1016/j.ppnp.2008.12.001} {\bibfield  {journal} {\bibinfo  {journal} {Prog. Part. Nucl. Phys.}\ }\textbf {\bibinfo {volume} {63}},\ \bibinfo {pages} {117 } (\bibinfo {year} {2009})}\BibitemShut {NoStop}%
\bibitem [{\citenamefont {Furnstahl}\ \emph {et~al.}(2015)\citenamefont {Furnstahl}, \citenamefont {Klco}, \citenamefont {Phillips},\ and\ \citenamefont {Wesolowski}}]{furnstahl2015}%
  \BibitemOpen
  \bibfield  {author} {\bibinfo {author} {\bibfnamefont {R.~J.}\ \bibnamefont {Furnstahl}}, \bibinfo {author} {\bibfnamefont {N.}~\bibnamefont {Klco}}, \bibinfo {author} {\bibfnamefont {D.~R.}\ \bibnamefont {Phillips}},\ and\ \bibinfo {author} {\bibfnamefont {S.}~\bibnamefont {Wesolowski}},\ }\bibfield  {title} {\bibinfo {title} {Quantifying truncation errors in effective field theory},\ }\href {https://doi.org/10.1103/PhysRevC.92.024005} {\bibfield  {journal} {\bibinfo  {journal} {Phys. Rev. C}\ }\textbf {\bibinfo {volume} {92}},\ \bibinfo {pages} {024005} (\bibinfo {year} {2015})}\BibitemShut {NoStop}%
\bibitem [{\citenamefont {Mishra}\ \emph {et~al.}(2022)\citenamefont {Mishra}, \citenamefont {Ekstr\"om}, \citenamefont {Hagen}, \citenamefont {Papenbrock},\ and\ \citenamefont {Platter}}]{mishra2022}%
  \BibitemOpen
  \bibfield  {author} {\bibinfo {author} {\bibfnamefont {C.}~\bibnamefont {Mishra}}, \bibinfo {author} {\bibfnamefont {A.}~\bibnamefont {Ekstr\"om}}, \bibinfo {author} {\bibfnamefont {G.}~\bibnamefont {Hagen}}, \bibinfo {author} {\bibfnamefont {T.}~\bibnamefont {Papenbrock}},\ and\ \bibinfo {author} {\bibfnamefont {L.}~\bibnamefont {Platter}},\ }\bibfield  {title} {\bibinfo {title} {Two-pion exchange as a leading-order contribution in chiral effective field theory},\ }\href {https://doi.org/10.1103/PhysRevC.106.024004} {\bibfield  {journal} {\bibinfo  {journal} {Phys. Rev. C}\ }\textbf {\bibinfo {volume} {106}},\ \bibinfo {pages} {024004} (\bibinfo {year} {2022})}\BibitemShut {NoStop}%
\end{thebibliography}%

\cleardoublepage
\section*{End Matter}
\label{sec:end_matter}

\textit{Details about the Hamiltonians.---}
The Hamiltonians employed in this work are described in detail in Refs.~\cite{elhatisari2016,elhatisari2017,lu2019}. 
We list the relevant expressions here for completeness.  
In what follows
\begin{equation}
\label{rowvec}
    a^\dagger(\mathbf{n}) \equiv \left[a_1^\dagger(\mathbf{n}),a_2^\dagger(\mathbf{n}),a_3^\dagger(\mathbf{n}),a_4^\dagger(\mathbf{n})\right]
\end{equation}
is a four component spin-isospin row vector. 
The kinetic energy is 
\begin{align}
	T &= -\frac{\hbar^2}{2ma^2}\sum\limits_\mathbf{n}\bigg[\frac{49}{6}a^\dagger(\mathbf{n})a(\mathbf{n}) - \frac{3}{2}\sum\limits_{|\mathbf{n}-\mathbf{n'}|=1}a^\dagger(\mathbf{n'})a(\mathbf{n})\nonumber\\
	&+\frac{3}{20}\sum\limits_{|\mathbf{n}-\mathbf{n'}|=2}a^\dagger(\mathbf{n'})a(\mathbf{n})- \frac{1}{90}\sum\limits_{|\mathbf{n}-\mathbf{n'}|=3}a^\dagger(\mathbf{n'})a(\mathbf{n})\bigg].
    \label{Tkin}
\end{align}

The one-pion exchange potential is
\begin{equation}
\label{ope}
	V_{\text{OPE}} = -\frac{g_A^2}{8\pi f_\pi^2}\sum\limits_{n',n,S',S,I}:\rho_{S',I}(\mathbf{n'})f_{S',S}(\mathbf{n'}-\mathbf{n})\rho_{S,I}(\mathbf{n}): .
\end{equation}
Here the spin-isospin density $\rho_{S,I}$ is defined for spin and isospin indices $S=x, y, z$ and $I=x, y, z$, respectively. Using  the Pauli spin matrices $\sigma_S$ and isospin matrices $\tau_I$ we have
\begin{equation}
\label{rhoSI}
	\rho_{S,I}(\mathbf{n}) = a^\dagger(\mathbf{n})\left[\sigma_S\otimes\tau_I\right]a(\mathbf{n}) \ ,
\end{equation}
and $f_{S',S}$ is given by
\begin{equation}
	f_{S',S}(\mathbf{n'}-\mathbf{n}) = \frac{1}{L^3}\sum\limits_{\mathbf{q}}\frac{e^{-i \mathbf{q}\dot(\mathbf{n'}-\mathbf{n})-b_\pi \mathbf{q}^2}\mathbf{q}_{S'}\mathbf{q}_S}{\mathbf{q}^2+m_\pi^2} \ .
\end{equation}
Here $\mathbf{q}_S$ is the $S$ component of the momentum transfer. It is an integer multiplied by $2\pi/L$.
$b_\pi$ acts like a cutoff for the interaction and uses the value $b_\pi = 0.7$. 
The constants used are the pion mass $m_\pi=134.98$ MeV, the axial-vector coupling constant $g_A=1.287$, and the pion decay constant $f_\pi=92.2$ MeV. The potential~(\ref{ope}) is in lattice units and needs to be multiplied with three powers of $\hbar/a=100$~MeV in actual computations.

In addition to the kinetic energy and the one-pion exchange, the Hamiltonian $H_B$ of Ref.~\cite{elhatisari2016} is consists of a combination of local and nonlocal potentials, see Eq.~(\ref{hamB}). 
The local potential is
\begin{align}
\label{VL}
    V_L &= \frac{c_{L}}{2}\sum_\mathbf{n} : {\rho}_L(\mathbf{n}) {\rho}_L(\mathbf{n}) : \nonumber\\
    &+\frac{c_{I,L}}{2}\sum_{I,\mathbf{n}} : {\rho}_{I,L}(\mathbf{n}) {\rho}_{I,L}(\mathbf{n}) :\notag\\
    &+\frac{c_{S,L}}{2}\sum_{S,\mathbf{n}} : {\rho}_{S,L}(\mathbf{n}) {\rho}_{S,L}(\mathbf{n}) : \notag\\
    &+ \frac{c_{S,I,L}}{2}\sum_{I,\mathbf{n}} : {\rho}_{S,I,L}(\mathbf{n}) {\rho}_{S,I,L}(\mathbf{n}) : .
\end{align}
Here we used the smeared local densities 
\begin{align}
    {\rho}_{L}(\mathbf{n}) &= {\rho}(\mathbf{n}) +s_L\sum_{|\mathbf{m}-\mathbf{n}|=1}{\rho}(\mathbf{m}) \notag\\
    {\rho}_{I,L}(\mathbf{n}) &= {\rho}_{I}(\mathbf{n}) +s_L\sum_{|\mathbf{m}-\mathbf{n}|=1}{\rho}_{I}(\mathbf{m}) \notag\\
    {\rho}_{S,L}(\mathbf{n}) &= {\rho}_{S}(\mathbf{n}) +s_L\sum_{|\mathbf{m}-\mathbf{n}|=1}{\rho}_{S}(\mathbf{m}) \notag\\
    {\rho}_{S,I,L}(\mathbf{n}) &= {\rho}_{S,I}(\mathbf{n}) +s_L\sum_{|\mathbf{m}-\mathbf{n}|=1}{\rho}_{S,I}(\mathbf{m}) \ , 
\end{align}
which are defined in terms of the local densities
\begin{align}
    {\rho}(\mathbf{n}) &= {a}^\dagger(\mathbf{n})a(\mathbf{n})\notag\\
    {\rho}_{I}(\mathbf{n}) &= {a}^\dagger(\mathbf{n})\left[\tau_I\right]{a}(\mathbf{n})\notag\\
    {\rho}_{S}(\mathbf{n}) &= {a}^\dagger(\mathbf{n})\left[\sigma_S\right]a(\mathbf{n}) \ ,
\end{align}
and $\rho_{S,I}(\mathbf{n})$ was already defined in Eq.~(\ref{rhoSI}). 

We turn to the nonlocal potential. 
Similarly to Eq.~(\ref{rowvec}) we combine the four components of the smeared operator~(\ref{tilde-a}) into a row vector 
\begin{equation}
    \tilde{a}^\dagger(\mathbf{n}) \equiv \left[\tilde{a}^\dagger_1(\mathbf{n}),\tilde{a}^\dagger_2(\mathbf{n}),\tilde{a}^\dagger_3(\mathbf{n}),\tilde{a}^\dagger_4(\mathbf{n})\right] \ .
\end{equation}
One defines the nonlocal density
\begin{equation}
\label{rhoNL}
    {\rho}_{NL}(\mathbf{n}) = \tilde{a}^\dagger(\mathbf{n})\tilde{a}(\mathbf{n}) \ ,
\end{equation}
and the nonlocal isospin density
\begin{equation}
\label{rhoINL}
    {\rho}_{I,NL}(\mathbf{n}) = \tilde{a}^\dagger(\mathbf{n})\left[\tau_I\right]\tilde{a}(\mathbf{n}) \ .
\end{equation}
The nonlocal potential then becomes
\begin{align}
    V_{NL} &= \frac{c_{NL}}{2}\sum_\mathbf{n} : {\rho}_{NL}(\mathbf{n}) {\rho}_{NL}(\mathbf{n}) : \notag\\
    &+ \frac{c_{I,NL}}{2}\sum_{I,\mathbf{n}} :{\rho}_{I,NL}(\mathbf{n}) {\rho}_{I,NL}(\mathbf{n}) : .
    \label{eq:VNL}
\end{align}

We see that the local potential $V_L$ only has local smearing, and that the nonlocal potential $V_{NL}$ only has nonlocal smearing. According to Ref.~\cite{elhatisari2016} the parameters of these potentials are 
$s_{NL} = 0.077$, $s_L = 0.81$, $c_{NL} = -0.1171$, $c_{I,NL} = 0.02607$, $c_L = -0.01013$, and $c_{S,L} = c_{I,L} = c_{S,I,L} = -c_L/3$. They were ``determined by
fitting to the low-energy nucleon-nucleon phase shifts, the observed deuteron energy, and the low-energy alpha-alpha
$S$-wave phase shifts.'' The results shown in Table~\ref{tab:res_combo} indicate that the alpha-alpha
$S$-wave phase shifts were not computed accurately. We also note that the significant local smearing $s_L=0.81$ implies that two nucleons interact strongly even when they are two lattice sites (i.e., almost 4~fm) apart. In numerical implementations one needs to multiply the couplings $c_{NL}$, $c_{I,NL}$, and $c_L$ by one power of $\hbar/a=100$~MeV.

\textit{Analytical expressions for the fully occupied lattice.---}
From Eq.~(\ref{Tkin}) we immediately see that the kinetic energy per particle is
\begin{equation}
\label{Tkin_full}
    \frac{\langle T \rangle}{A} = \frac{49}{12ma^2} \ .  
\end{equation}
A somewhat lengthy derivation [see Supplemental Material for details] yields the expectation value of the two-body potential~(\ref{cont}) on a fully occupied lattice with spin-isospin degeneracy $g$. The result is
\begin{align}
\label{full}
    \frac{\langle V_0 \rangle}{c_0A} &= \frac{1}{2g}\left\langle :\tilde{\rho}^2(\mathbf{n}):\right\rangle
    \nonumber\\
    &=\left[\frac{g-1}{2}+6gs_L+3(6g-1)s_L^2\right] \left(1+6s_{NL}^2\right)^2\nonumber\\
    &-3s_Ls_{NL}^2\left(8+17s_Ls_{NL}^2\right)  \ .
\end{align}
Similarly, we obtain the following expression for the three-body potential.
\begin{align}
\label{Wfull}
    \frac{\langle W \rangle}{c_3 A} &= \frac{1}{3!g}\left\langle :\tilde{\rho}^3(\mathbf{n}):\right\rangle\nonumber\\
    &=\left(1+6s_{NL}^2\right)^3\Big\{\tfrac{1}{6}(g-1)(g-2)+3g(g-1)s_L \nonumber\\
    &+3g(6g-1)s_L^2+2\left[1+9g(2g-1)\right]s_L^3 \Big\}\nonumber\\
    &-\left(1+6s_{NL}^2\right)\Big[24(g-1)s_Ls_{NL}^2+24(6g-1)s_L^2s_{NL}^2 \nonumber\\
    &+51gs_L^2s_{NL}^4 + 102(3g-1)s_L^3s_{NL}^4\Big]\nonumber\\
    &+8s_L^2s_{NL}^4\left(27+28s_Ls_{NL}\right)  \ .
\end{align}

To verify our nuclear matter results, we compute the energy expectation value of a completely filled lattice, i.e., for $A=4L^3$.
Results are presented in Table~\ref{tab:full} for the Hamiltonian~(\ref{ham_chiral}). The expectation value $\langle T+V_0\rangle/A$ is close to the full data points shown at the maximum density in Fig.~\ref{fig:EOS}. We also see that the one-pion exchange acts repulsive and contributes only about $+1.25$~MeV per nucleon.

\begin{table}[t!]
\renewcommand{\arraystretch}{1.2}
\centering
\caption{Kinetic and potential energies per nucleon (in MeV) for a fully filled lattice for the potentials of the Hamiltonian~(\ref{ham_chiral}) from Ref.~\cite{elhatisari2017}.}
\label{tab:full}
\renewcommand{\arraystretch}{1.3}
\begin{ruledtabular}
\begin{tabular}{crr}
$V$ & $\langle V\rangle/A$ & $\langle T+V\rangle/A$ \\
\hline
$V_0$ & $-76.80$ & $-33.31$ \\
$V_0+V_{\rm OPE}$& $-75.55$ & $-32.06$
\end{tabular}
\end{ruledtabular}
\end{table}

\textit{Continuum limit.---}
To address the continuum limit of vanishing lattice spacing $a\to 0$, we focus on low densities. There, the two-body potential dominates (see Fig.~\ref{fig:EOS-contrib}), and at the lowest-density point $\rho=4/(aL)^3$ one finds analytically
\begin{equation}
\label{lowdens}
    \frac{\langle V_0 \rangle}{A} = \tfrac{3}{8}c_0 a^3 \left(1+6s_{L}\right)^2\left(1+6s_{NL}\right)^4 \rho \ .
\end{equation}
We use this density for renormalization (i.e., for adjusting $c_0$ at $\rho=4/(aL)^3$ such that $\langle V_0\rangle/A$ is fixed) and take $a\to 0$, $L\to\infty$ while keeping the volume $(aL)^3$ fixed.  This requires $c_0\propto a^{-3}$. Using Eqs.~(\ref{full}) and (\ref{Tkin_full}) this then implies that $-\langle V_0\rangle/A \propto a^{-3} \gg \langle T\rangle/A\propto a^{-2}$ when $a\to 0$, $L\to\infty$ and $aL=\mathrm{const}$ for the fully occupied lattice. Thus, in the continuum limit the two-body interaction~(\ref{cont}) leads to a collapse of the system. Alternatively, one might renormalize at low densities by keeping $c_0$ fixed and by scaling $s_{NL}\propto a^{-1/2}$ and $s_{L}\propto a^{-1/2}$ as $a\to 0$. For the fully occupied lattice one then again finds  from Eq.~(\ref{full}) that $\langle V_0\rangle/A \propto a^{-3}$, which leads to collapse.

When using nucleons as the degrees of freedom there are limited reasons why one would want to decrease the lattice spacing beyond the breakdown spacing $a \approx \pi/\Lambda_\chi$ where $\Lambda_\chi\approx 0.7$~GeV is the chiral breakdown scale~\cite{furnstahl2015,mishra2022}. Thus, the continuum limit is not of too much interest in nuclear lattice effective field theory. However, our analysis makes clear that the lattice potential of Ref.~\cite{lu2019} cannot simply be used in computations that employ continuum Hamiltonians.   

\cleardoublepage
\onecolumngrid
\begin{center}
\textbf{Nuclear Saturation from Lattice Hamiltonians: \\Supplemental Material}
\end{center}
\twocolumngrid

\section{Reference states for nuclei}
We start our Hartree-Fock computations for nuclei from compact initial states.
For $^4$He, our initial state is chosen to be four nucleons at the origin
$\mathbf{n}=(0,0,0)$,
two neutrons with spins up and down
and two protons with spins up and down.
The lattice Hamiltonians we explore are translationally invariant, so this gives the same energy as defining our initial state at any other site.
For $^8$Be,
we choose the initial state to be four nucleons at
$\mathbf{n}=(0,0,0)$
and four nucleons at
$\mathbf{n}=(1,0,0)$.
Other possible compact initial states are equivalent up to translations and cubic rotations.

For $^{12}$C, the obvious choice is a ``right-angled'' configuration:
four nucleons at each of
$\mathbf{n}=(0,0,0)$,
$(1,0,0)$,
and $(0,1,0)$.
However, other low-lying configurations are also possible.
We explore these in Table~\ref{tab:InitialStates}.
The right-angled configuration is clearly lower in energy than the linear configuration
or the ``bent-arm'' configuration.
In smaller lattices,
we occasionally find that the linear configuration gives us the lowest-energy Hartree-Fock state
due to large finite-volume effects.
This vanishes as we go to large lattices.
For $^{16}$O,
we also find that the ``right-angled'' configuration,
with four nucleons at each of
$\mathbf{n}=(0,0,0)$,
$(1,0,0)$,
$(0,1,0)$,
and $(0,0,1)$,
gives the lowest Hartree-Fock energy.

\begin{table}[tb!]
\renewcommand{\arraystretch}{1.2}
\centering
\caption{Hartree-Fock energies in MeV for different initial states for the Hamiltonians $H$~\cite{elhatisari2017} and $H_B$~\cite{elhatisari2016} for lattice extent $L=6$.
For each configuration, we give the lattice sites at which four nucleons are placed to create the initial state.}
\label{tab:InitialStates}
\renewcommand{\arraystretch}{1.3}
\begin{ruledtabular}
\begin{tabular}{l c d d}
Nucl. & Configuration & \multicolumn{1}{c}{$H$} & \multicolumn{1}{c}{$H_B$}\\
\hline
$^{4}$He & $(0,0,0)$ & -19.81 & -32.19\\
\hline
$^{8}$Be & $(0,0,0), (1,0,0)$ & -63.17 & -76.37\\
\hline
\multirow{3}{*}{$^{12}$C} & $(0,0,0), (1,0,0), (0,1,0)$ & -137.28 & -139.83\\
& $(0,0,0), (1,0,0), (2,0,0)$ & -107.64 & -128.12\\
& $(0,0,0), (1,0,0), (2,1,0)$ & -101.26 & -126.72 \\
\hline
\multirow{4}{*}{$^{16}$O} & $(0,0,0), (1,0,0), (0,1,0), (0,0,1)$ & -213.37 & -222.06 \\
& $(0,0,0), (1,0,0), (1,1,0), (0,1,0)$ & -187.22 & -212.84 \\
& $(0,0,0), (1,0,0), (1,1,0), (2,0,0)$ & -192.07 & -210.72 \\
& $(0,0,0), (1,0,0), (2,0,0), (3,0,0)$ & -151.51 & -180.94
\end{tabular}
\end{ruledtabular}
\end{table}

\section{Two-nucleon benchmarks}
For the Hamiltonian~(\ref{ham_chiral}) of Ref.~\cite{elhatisari2017} we list results from exact diagonalization for the deuteron in Table~\ref{tab:bench_d} and for the two-neutron system in Table~\ref{tab:bench_nn}. When the one-pion exchange is neglected, two neutrons and the deuteron have the same energy, because $V_0$ is spin-isospin symmetric. We see that the one-pion exchange is very weak and contributes little to the ground-state energy. 

\begin{table}[t!]
\renewcommand{\arraystretch}{1.2}
\centering
\caption{Ground-state energies (in MeV) of the deuteron for the potentials and lattice sizes as indicated, obtained from exact diagonalization. Parameters are from Ref.~\cite{elhatisari2017}.}
\label{tab:bench_d}
\renewcommand{\arraystretch}{1.3}
\begin{ruledtabular}
\begin{tabular}{clll}
$L$ & $V_0 +V_{\rm OPE}$ & $V_0$    & $V_{\rm OPE}$   \\
\hline
4  & $-5.04041823824$ &  $-5.17260862889$  & $-0.0132647363266$\\
5  & $-3.26319416528$ &  $-3.42452503725$  & $-0.0139277520259$\\
6  & $-2.38697021090$ &  $-2.55984789719$  & $-0.0107662557934$\\
7  & $-1.88890653391$ &  $-2.06746886103$  & $-0.0076071101145$\\
\end{tabular}
\end{ruledtabular}
\end{table}

\begin{table}[t!]
\renewcommand{\arraystretch}{1.2}
\centering
\caption{Same as Table~\ref{tab:bench_d} but for the two-neutron system.}
\label{tab:bench_nn}
\renewcommand{\arraystretch}{1.3}
\begin{ruledtabular}
\begin{tabular}{cll}
$L$ & \multicolumn{1}{c}{$V_0 +V_{\rm OPE}$} & \multicolumn{1}{c}{$V_{\rm OPE}$} \\
\hline
4  &	$-5.03104190761$ &  $-0.00147403310330$  \\
5  &	$-3.25336204693$ &  $-0.00154796654782$  \\
6  &    $-2.37865887117$ &  $-0.00119657968867$  \\
7  &    $-1.88190245681$ &  $-0.00084529617778$  \\
\end{tabular}
\end{ruledtabular}
\end{table}

\section{Expectation values in the fully occupied lattice}
\label{full_derivation}
The interaction of Refs.~\cite{elhatisari2017} contains a smeared ``contact'' as the dominant term (the one-pion exchange is very weak). It is based on the smeared creation operator~(\ref{tilde-a}) 
and the smeared density [see Eq.~(\ref{tilderho})]
\begin{equation}
\label{densNL}
    \tilde{\rho}(\mathbf{n}) =\sum_{i=1}^g  \left(\tilde{a}_i^\dagger(\mathbf{n}) \tilde{a}_i(\mathbf{n}) + s_L\sum_{|\mathbf{m}-\mathbf{n}|=1}\tilde{a}_i^\dagger(\mathbf{m})\tilde{a}_i(\mathbf{m}) \right).
\end{equation}
Here we introduced the spin-isospin degeneracy $g$. 

We want to compute the expectation values of $:\tilde{\rho}^2(\mathbf{n}):$ and $:\tilde{\rho}^3(\mathbf{n}):$ in the fully occupied lattice. To do so, we note that the smeared creation operator~(\ref{tilde-a}) and its adjoint fulfill the anti-commutation relation
\begin{equation}
\label{comm}
    \left\{\tilde{a}_i(\mathbf{n}),  \tilde{a}_j^\dagger(\mathbf{n}) \right\}= \left(1+6s_{NL}^2\right)\delta_{ij} \ . 
\end{equation}
The smeared density consists of smeared creation and annihilation operators. Let $\mathbf{m}$ and $\mathbf{m}'$ be nearest neighbors of $\mathbf{n}$. We have
\begin{equation}
\label{comm_mn}
    \left\{\tilde{a}_i(\mathbf{n}),  \tilde{a}_j^\dagger(\mathbf{m}) \right\}= 2s_{NL}\delta_{ij} \ , 
\end{equation}
and 
\begin{align}
\label{comm_mm}
\left\{\tilde{a}_i(\mathbf{m}),  \tilde{a}_j^\dagger(\mathbf{m}') \right\}=
\delta_{ij}\left\{\begin{array}{cl}
     1+6s_{NL}^2 &  \mbox{if $\mathbf{m}=\mathbf{m}'$,}\\
     s_{NL}^2 &  \mbox{if $\mathbf{m}=-\mathbf{m}'$,}\\
     2s_{NL}^2 &  \mbox{else.} 
\end{array}\right.
\end{align}
Here, $\mathbf{m}=-\mathbf{m}'$ is understood such that $\mathbf{m}$ and $\mathbf{m'}$ are on opposite sites of $\mathbf{n}$. For fixed $\mathbf{n}$, there are six possible sites for each $\mathbf{m}$ and $\mathbf{m'}$. For fixed $\mathbf{n}$ and $\mathbf{m}$ the three possible values of the anti-commutator~(\ref{comm_mm}) occur with frequency one, one, and four when going from top to bottom on the right-hand-side of Eq.~(\ref{comm_mm}).

We are now ready to evaluate the expectation values of $:\tilde{\rho}^2(\mathbf{n}):$ and $:\tilde{\rho}^3(\mathbf{n}):$ in the fully occupied lattice. Starting with Eq.~(\ref{densNL}) we have
\begin{align}
    :\tilde{\rho}^2(\mathbf{n}): &= \sum_{i,j=1}^g \tilde{a}_i^\dagger(\mathbf{n})  \tilde{a}_j^\dagger(\mathbf{n})
    \tilde{a}_j(\mathbf{n})  \tilde{a}_i(\mathbf{n})\nonumber\\
    &+2s_L\sum_{\mathbf{m}}\sum_{i,j=1}^g \tilde{a}_i^\dagger(\mathbf{n})  \tilde{a}_j^\dagger(\mathbf{m})
    \tilde{a}_j(\mathbf{m})  \tilde{a}_i(\mathbf{n})\nonumber\\
    &+s_L^2\sum_{\mathbf{m},\mathbf{m'}}\sum_{i,j=1}^g \tilde{a}_i^\dagger(\mathbf{m})  \tilde{a}_j^\dagger(\mathbf{m}')
    \tilde{a}_j(\mathbf{m}')  \tilde{a}_i(\mathbf{m}) \ .
\end{align}
Evaluating the expectation value in the fully occupied lattice requires us to perfrom Wick contractions using the anti-commutation relations~(\ref{comm}), (\ref{comm_mn}), and (\ref{comm_mm}). 
Clearly, if we move the creation operators to the right, they will annihilate the fully occupied lattice. We have
\begin{align}
    \sum_{i,j=1}^g 
    \contraction[2ex]{}{\tilde{a}}
    {_i^\dagger(\mathbf{n})  \tilde{a}_j^\dagger(\mathbf{n})
    \tilde{a}_j(\mathbf{n})}{\tilde{a}}
\contraction[1ex]{\tilde{a}_i^\dagger(\mathbf{n})}{\tilde{a}}{_j^\dagger(\mathbf{n})}{\tilde{a}}
    \tilde{a}_i^\dagger(\mathbf{n})  \tilde{a}_j^\dagger(\mathbf{n})
    \tilde{a}_j(\mathbf{n})  \tilde{a}_i(\mathbf{n}) \to g^2\left(1+6s_{NL}^2\right)^2 \ , 
\end{align}
and
\begin{align}
\label{special1}
    \sum_{i,j=1}^g 
    \contraction{}
    {\tilde{a}}
    {_i^\dagger(\mathbf{n})\tilde{a}_j(\mathbf{n})}
    {\tilde{a}}
\contraction[2ex]{\tilde{a}_i^\dagger(\mathbf{n})}{\tilde{a}}
{_j^\dagger(\mathbf{n})
    \tilde{a}_j(\mathbf{n})}
{\tilde{a}}
    \tilde{a}_i^\dagger(\mathbf{n})  \tilde{a}_j^\dagger(\mathbf{n})
    \tilde{a}_j(\mathbf{n})  \tilde{a}_i(\mathbf{n}) \to -g\left(1+6s_{NL}^2\right)^2 \ , 
\end{align}
and
\begin{align}
    \sum_{\mathbf{m}}\sum_{i,j=1}^g 
    \contraction[2ex]{}{\tilde{a}}
    {_i^\dagger(\mathbf{n})  \tilde{a}_j^\dagger(\mathbf{m})
    \tilde{a}_j(\mathbf{m})}{\tilde{a}}
\contraction[1ex]{\tilde{a}_i^\dagger(\mathbf{n})}{\tilde{a}}{_j^\dagger(\mathbf{m})}{\tilde{a}}
    \tilde{a}_i^\dagger(\mathbf{n})  \tilde{a}_j^\dagger(\mathbf{m})
    \tilde{a}_j(\mathbf{m})  \tilde{a}_i(\mathbf{n}) \to 6g^2\left(1+6s_{NL}^2\right)^2 \ , 
\end{align}
and
\begin{align}
\label{special2}
    \sum_{\mathbf{m}}\sum_{i,j=1}^g 
    \contraction{}
    {\tilde{a}}
    {_i^\dagger(\mathbf{n})\tilde{a}_j(\mathbf{m})}
    {\tilde{a}}
\contraction[2ex]{\tilde{a}_i^\dagger(\mathbf{n})}{\tilde{a}}
{_j^\dagger(\mathbf{m})
    \tilde{a}_j(\mathbf{m})}
{\tilde{a}}
    \tilde{a}_i^\dagger(\mathbf{n})  \tilde{a}_j^\dagger(\mathbf{m})
    \tilde{a}_j(\mathbf{m})  \tilde{a}_i(\mathbf{n}) \to -6g\left(2s_{NL}\right)^2 \ , 
\end{align}
and
\begin{align}
    \sum_{\mathbf{m},\mathbf{m'}}&\sum_{i,j=1}^g 
    \contraction[2ex]{}{\tilde{a}}
    {_i^\dagger(\mathbf{m})  \tilde{a}_j^\dagger(\mathbf{m}')
    \tilde{a}_j(\mathbf{m}')}{\tilde{a}}
\contraction[1ex]{\tilde{a}_i^\dagger(\mathbf{m})}{\tilde{a}}{_j^\dagger(\mathbf{m}')}{\tilde{a}}
    \tilde{a}_i^\dagger(\mathbf{m})  \tilde{a}_j^\dagger(\mathbf{m}')
    \tilde{a}_j(\mathbf{m}')  \tilde{a}_i(\mathbf{m}) \nonumber\\
    \to& \,\, (6g)^2\left(1+6s_{NL}^2\right)^2 \ , 
\end{align}
and
\begin{align}
\label{special3}
    \sum_{\mathbf{m},\mathbf{m'}}&\sum_{i,j=1}^g 
    \contraction{}
    {\tilde{a}}
    {_i^\dagger(\mathbf{m})\tilde{a}_j(\mathbf{m}')}
    {\tilde{a}}
\contraction[2ex]{\tilde{a}_i^\dagger(\mathbf{m})}{\tilde{a}}
{_j^\dagger(\mathbf{m}')
    \tilde{a}_j(\mathbf{m}')}
{\tilde{a}}
    \tilde{a}_i^\dagger(\mathbf{m})  \tilde{a}_j^\dagger(\mathbf{m}')
    \tilde{a}_j(\mathbf{m}')  \tilde{a}_i(\mathbf{m}) \nonumber\\
    \to& -6g\left[\left(1+6s_{NL}^2\right)^2 + s_{NL}^4 + 4(2s_{NL}^2)^2 \right] \nonumber\\
    =& -6g\left[\left(1+6s_{NL}^2\right)^2 + 17s_{NL}^4 \right] \ . 
\end{align}
Thus, we obtain for the expectation value of $:\tilde{\rho}^2(\mathbf{n})/2:$ per nucleon  
\begin{align}
    \frac{1}{2g}\left\langle :\tilde{\rho}^2(\mathbf{n}):\right\rangle &=
    \frac{g-1}{2}\left(1+6s_{NL}^2\right)^2 \nonumber\\
    &+6s_L\left[g\left(1+6s_{NL}^2\right)^2-4s_{NL}^2\right]\nonumber\\
    &+3s_L^2\left[(6g-1)\left(1+6s_{NL}^2\right)^2-17s_{NL}^4\right]\ .
\end{align}
This is essentially, i.e., up to the factor $c_0$, the expression~(\ref{full}).

We turn to $:\tilde{\rho}^3(\mathbf{n}):$ and expand
\begin{align}
\label{rho3}
    :&\tilde{\rho}^3(\mathbf{n}): = \sum_{ijk} \tilde{a}_i^\dagger(\mathbf{n})  \tilde{a}_j^\dagger(\mathbf{n})
    \tilde{a}_k^\dagger(\mathbf{n})
    \tilde{a}_k(\mathbf{n}) 
    \tilde{a}_j(\mathbf{n})  \tilde{a}_i(\mathbf{n})\nonumber\\
    &+3s_L\sum_{\mathbf{m}}\sum_{ijk} \tilde{a}_i^\dagger(\mathbf{n})  \tilde{a}_j^\dagger(\mathbf{n})
    \tilde{a}_k^\dagger(\mathbf{m})
    \tilde{a}_k(\mathbf{m}) 
    \tilde{a}_j(\mathbf{n})  \tilde{a}_i(\mathbf{n})\nonumber\\
    &+3s_L^2\sum_{\mathbf{m},\mathbf{m'}}\sum_{ijk} \tilde{a}_i^\dagger(\mathbf{n}) \tilde{a}_j^\dagger(\mathbf{m})  \tilde{a}_k^\dagger(\mathbf{m}')
    \tilde{a}_k(\mathbf{m}')  \tilde{a}_j(\mathbf{m}) 
    \tilde{a}_i(\mathbf{n}) \nonumber\\
    &+s_L^3\sum_{\mathbf{l},\mathbf{m},\mathbf{m'}}\sum_{ijk} \tilde{a}_i^\dagger(\mathbf{m}) \tilde{a}_j^\dagger(\mathbf{m'})  \tilde{a}_k^\dagger(\mathbf{l})
    \tilde{a}_k(\mathbf{l})  \tilde{a}_j(\mathbf{m}') 
    \tilde{a}_i(\mathbf{m})\ .
\end{align}
Here, it is understood that $\mathbf{l}$ also is a nearest neighbor of $\mathbf{n}$. There are six different ways to fully contract each line on the right-hand side of Eq.~(\ref{rho3}). The contractions of the first line of the right-hand side of Eq.~(\ref{rho3}) are standard and yield
\begin{equation}
    g(g-1)(g-2)\left(1+6s_{NL}^2\right)^3 \ .
\end{equation}
For the second line of the right-hand side of Eq.~(\ref{rho3}) we find
\begin{align}
\sum_{\mathbf{m}}&\sum_{ijk}
\contraction[3ex]{}
{\tilde{a}}
{_i^\dagger(\mathbf{n})  \tilde{a}_j^\dagger(\mathbf{n})
\tilde{a}_k^\dagger(\mathbf{m})
\tilde{a}_k(\mathbf{m}) 
\tilde{a}_j(\mathbf{n}}
{\tilde{a}}
\contraction[2ex]
{\tilde{a}_i^\dagger(\mathbf{n})}
{\tilde{a}}
{_j^\dagger(\mathbf{n})
\tilde{a}_k^\dagger(\mathbf{m})
\tilde{a}_k(\mathbf{m})}
{\tilde{a}}
\contraction[1ex]
{\tilde{a}_i^\dagger(\mathbf{n})  \tilde{a}_j^\dagger(\mathbf{n})}
{\tilde{a}}
{_k^\dagger(\mathbf{m})}
{\tilde{a}}
\tilde{a}_i^\dagger(\mathbf{n})  \tilde{a}_j^\dagger(\mathbf{n})
\tilde{a}_k^\dagger(\mathbf{m})
\tilde{a}_k(\mathbf{m}) 
\tilde{a}_j(\mathbf{n})
\tilde{a}_i(\mathbf{n}) \nonumber\\
\to& \,\, 6 g^3 \left(1+6s_{NL}^2\right)^3 \ ,
\end{align}
and 
\begin{align}
\sum_{\mathbf{m}}&\sum_{ijk}
\contraction[2ex]{}
{\tilde{a}}
{_i^\dagger(\mathbf{n})  \tilde{a}_j^\dagger(\mathbf{n})
\tilde{a}_k^\dagger(\mathbf{m})
\tilde{a}_k(\mathbf{m})}
{\tilde{a}}
\contraction[3ex]{\tilde{a}_i^\dagger(\mathbf{n})}
{\tilde{a}}
{_j^\dagger(\mathbf{n})
\tilde{a}_k^\dagger(\mathbf{m})
\tilde{a}_k(\mathbf{m}) 
\tilde{a}_j(\mathbf{n})}
{\tilde{a}}
\contraction[1ex]
{\tilde{a}_i^\dagger(\mathbf{n})  \tilde{a}_j^\dagger(\mathbf{n})}
{\tilde{a}}
{_k^\dagger(\mathbf{m})}
{\tilde{a}}
\tilde{a}_i^\dagger(\mathbf{n})  \tilde{a}_j^\dagger(\mathbf{n})
\tilde{a}_k^\dagger(\mathbf{m})
\tilde{a}_k(\mathbf{m}) 
\tilde{a}_j(\mathbf{n})
\tilde{a}_i(\mathbf{n}) \nonumber\\
\to& -6g^2\left(1+6s_{NL}^2\right)^3 \ , 
\end{align}
and 
\begin{align}
\sum_{\mathbf{m}}&\sum_{ijk}
\contraction[3ex]
{}
{\tilde{a}}
{_i^\dagger(\mathbf{n})  \tilde{a}_j^\dagger(\mathbf{n})
\tilde{a}_k^\dagger(\mathbf{m})
\tilde{a}_k(\mathbf{m}) 
\tilde{a}_j(\mathbf{n})}
{\tilde{a}}
\contraction[1ex]
{\tilde{a}_i^\dagger(\mathbf{n})  \tilde{a}_j^\dagger(\mathbf{n})}
{\tilde{a}}
{_k^\dagger(\mathbf{m})
\tilde{a}_k(\mathbf{m})}
{\tilde{a}}
\contraction[2ex]
{\tilde{a}_i^\dagger(\mathbf{n})}
{\tilde{a}}
{_j^\dagger(\mathbf{n})
\tilde{a}_k^\dagger(\mathbf{m})}
{\tilde{a}}
\tilde{a}_i^\dagger(\mathbf{n})  \tilde{a}_j^\dagger(\mathbf{n})
\tilde{a}_k^\dagger(\mathbf{m})
\tilde{a}_k(\mathbf{m}) 
\tilde{a}_j(\mathbf{n})
\tilde{a}_i(\mathbf{n}) \nonumber\\
\to& -2\times 6 g^2 \left(1+6s_{NL}^2\right)(2s_{NL})^2   \ , 
\end{align}
and 
\begin{align}
\sum_{\mathbf{m}}&\sum_{ijk}
\contraction[3ex]
{}
{\tilde{a}}
{_i^\dagger(\mathbf{n})  \tilde{a}_j^\dagger(\mathbf{n})
\tilde{a}_k^\dagger(\mathbf{m})}
{\tilde{a}}
\contraction[2ex]
{\tilde{a}_i^\dagger(\mathbf{n})}
{\tilde{a}}
{_j^\dagger(\mathbf{n})
\tilde{a}_k^\dagger(\mathbf{m})
\tilde{a}_k(\mathbf{m}) 
\tilde{a}_j(\mathbf{n})}
{\tilde{a}}
\contraction[1ex]
{\tilde{a}_i^\dagger(\mathbf{n})  \tilde{a}_j^\dagger(\mathbf{n})}
{\tilde{a}}
{_k^\dagger(\mathbf{m})
\tilde{a}_k(\mathbf{m})}
{\tilde{a}}
\tilde{a}_i^\dagger(\mathbf{n})  \tilde{a}_j^\dagger(\mathbf{n})
\tilde{a}_k^\dagger(\mathbf{m})
\tilde{a}_k(\mathbf{m}) 
\tilde{a}_j(\mathbf{n})
\tilde{a}_i(\mathbf{n}) \nonumber\\
\to& \,\, 2\times 6 g \left(1+6s_{NL}^2\right)(2s_{NL})^2 \ .   
\end{align}
The last two contractions come with factors of two, as clearly indicated, because there are two equivalent ways of performing the contractions (but only one is shown). Thus the four different contractions shown above capture the six different ways to perform them.

For the third line of the right-hand side of Eq.~(\ref{rho3}) we find
\begin{align}
\sum_{\mathbf{m},\mathbf{m'}}&\sum_{ijk}
\contraction[3ex]
{}
{\tilde{a}}
{_i^\dagger(\mathbf{n}) \tilde{a}_j^\dagger(\mathbf{m})  \tilde{a}_k^\dagger(\mathbf{m}')
\tilde{a}_k(\mathbf{m}')  
\tilde{a}_j(\mathbf{m})}
{\tilde{a}}
\contraction[2ex]
{\tilde{a}_i^\dagger(\mathbf{n})}
{\tilde{a}}
{_j^\dagger(\mathbf{m})  \tilde{a}_k^\dagger(\mathbf{m}')
\tilde{a}_k(\mathbf{m}')}
{\tilde{a}}
\contraction[1ex]
{\tilde{a}_i^\dagger(\mathbf{n}) \tilde{a}_j^\dagger(\mathbf{m})}
{\tilde{a}}
{_k^\dagger(\mathbf{m}')}
{\tilde{a}}
\tilde{a}_i^\dagger(\mathbf{n}) \tilde{a}_j^\dagger(\mathbf{m})  \tilde{a}_k^\dagger(\mathbf{m}')
\tilde{a}_k(\mathbf{m}')  
\tilde{a}_j(\mathbf{m}) 
\tilde{a}_i(\mathbf{n}) \nonumber\\
\to& \,\, 36g^3 \left(1+6s_{NL}^2\right)^3\ , 
\end{align}
and
\begin{align}
\sum_{\mathbf{m},\mathbf{m'}}&\sum_{ijk}
\contraction[3ex]
{}
{\tilde{a}}
{_i^\dagger(\mathbf{n}) \tilde{a}_j^\dagger(\mathbf{m})  \tilde{a}_k^\dagger(\mathbf{m}')
\tilde{a}_k(\mathbf{m}')  
\tilde{a}_j(\mathbf{m})}
{\tilde{a}}
\contraction[2ex]
{\tilde{a}_i^\dagger(\mathbf{n})}
{\tilde{a}}
{_j^\dagger(\mathbf{m})  \tilde{a}_k^\dagger(\mathbf{m}')}
{\tilde{a}}
\contraction[1ex]
{\tilde{a}_i^\dagger(\mathbf{n}) \tilde{a}_j^\dagger(\mathbf{m})}
{\tilde{a}}
{_k^\dagger(\mathbf{m}')
\tilde{a}_k(\mathbf{m}')}
{\tilde{a}}
\tilde{a}_i^\dagger(\mathbf{n}) \tilde{a}_j^\dagger(\mathbf{m})  \tilde{a}_k^\dagger(\mathbf{m}')
\tilde{a}_k(\mathbf{m}')  
\tilde{a}_j(\mathbf{m})
\tilde{a}_i(\mathbf{n}) \nonumber\\
\to&  -6g^2\left(1+6s_{NL}^2\right)\left[\left(1+6s_{NL}^2\right)^2 + 17s_{NL}^4\right] \ , 
\end{align}
and 
\begin{align}
\sum_{\mathbf{m},\mathbf{m'}}&\sum_{ijk}
\contraction[3ex]
{}
{\tilde{a}}
{_i^\dagger(\mathbf{n}) \tilde{a}_j^\dagger(\mathbf{m})  \tilde{a}_k^\dagger(\mathbf{m}')
\tilde{a}_k(\mathbf{m}')}
{\tilde{a}}
\contraction[2ex]
{\tilde{a}_i^\dagger(\mathbf{n})}
{\tilde{a}}
{_j^\dagger(\mathbf{m})  \tilde{a}_k^\dagger(\mathbf{m}')
\tilde{a}_k(\mathbf{m}')  
\tilde{a}_j(\mathbf{m})}
{\tilde{a}}
\contraction[1ex]
{\tilde{a}_i^\dagger(\mathbf{n}) \tilde{a}_j^\dagger(\mathbf{m})}
{\tilde{a}}
{_k^\dagger(\mathbf{m}')}
{\tilde{a}}
\tilde{a}_i^\dagger(\mathbf{n}) \tilde{a}_j^\dagger(\mathbf{m})  \tilde{a}_k^\dagger(\mathbf{m}')
\tilde{a}_k(\mathbf{m}')  
\tilde{a}_j(\mathbf{m})
\tilde{a}_i(\mathbf{n}) \nonumber\\
\to& - 2\times (6g)^2\left(1+6s_{NL}^2\right)(2s_{NL})^2
\end{align}
and
\begin{align}
\sum_{\mathbf{m},\mathbf{m'}}&\sum_{ijk}
\contraction[3ex]
{}
{\tilde{a}}
{_i^\dagger(\mathbf{n}) \tilde{a}_j^\dagger(\mathbf{m})  \tilde{a}_k^\dagger(\mathbf{m}')
\tilde{a}_k(\mathbf{m}')}
{\tilde{a}}
\contraction[2ex]
{\tilde{a}_i^\dagger(\mathbf{n})}
{\tilde{a}}
{_j^\dagger(\mathbf{m})  \tilde{a}_k^\dagger(\mathbf{m}')}
{\tilde{a}}
\contraction[1ex]
{\tilde{a}_i^\dagger(\mathbf{n}) \tilde{a}_j^\dagger(\mathbf{m})}
{\tilde{a}}
{_k^\dagger(\mathbf{m}')
\tilde{a}_k(\mathbf{m}')  
\tilde{a}_j(\mathbf{m})}
{\tilde{a}}
\tilde{a}_i^\dagger(\mathbf{n}) \tilde{a}_j^\dagger(\mathbf{m})  \tilde{a}_k^\dagger(\mathbf{m}')
\tilde{a}_k(\mathbf{m}')  
\tilde{a}_j(\mathbf{m}) 
\tilde{a}_i(\mathbf{n}) \nonumber\\
\to& \,\, 2\times 6g (2s_{NL})^2\left(1+15s_{NL}^2\right) \ .
\end{align}
As indicated, the last two contractions again come with factors of two. 

Finally, for the last line of the right-hand side of Eq.~(\ref{rho3}) we find
\begin{align}
\sum_{\mathbf{l},\mathbf{m},\mathbf{m'}}&\sum_{ijk}
\contraction[3ex]
{}
{\tilde{a}}
{_i^\dagger(\mathbf{m}) \tilde{a}_j^\dagger(\mathbf{m'})  \tilde{a}_k^\dagger(\mathbf{\nu})
\tilde{a}_k(\mathbf{l})
\tilde{a}_j(\mathbf{m}')}
{\tilde{a}}
\contraction[2ex]
{\tilde{a}_i^\dagger(\mathbf{m})}
{\tilde{a}}
{_j^\dagger(\mathbf{m'})  \tilde{a}_k^\dagger(\mathbf{\nu})
\tilde{a}_k(\mathbf{l})}
{\tilde{a}}
\contraction[1ex]
{\tilde{a}_i^\dagger(\mathbf{m}) \tilde{a}_j^\dagger(\mathbf{m'})}
{\tilde{a}}
{_k^\dagger(\mathbf{l})}
{\tilde{a}}
\tilde{a}_i^\dagger(\mathbf{m}) \tilde{a}_j^\dagger(\mathbf{m'})  \tilde{a}_k^\dagger(\mathbf{l})
\tilde{a}_k(\mathbf{l})
\tilde{a}_j(\mathbf{m}') 
\tilde{a}_i(\mathbf{m})    \nonumber\\
\to& \,\, (6g)^3 \left(1+6s_{NL}^2\right)^3 \ , 
\end{align}
and
\begin{align}
\sum_{\mathbf{l},\mathbf{m},\mathbf{m'}}&\sum_{ijk}
\contraction[3ex]
{}
{\tilde{a}}
{_i^\dagger(\mathbf{m}) \tilde{a}_j^\dagger(\mathbf{m'})  \tilde{a}_k^\dagger(\mathbf{l})
\tilde{a}_k(\mathbf{l})
\tilde{a}_j(\mathbf{m}')}
{\tilde{a}}
\contraction[2ex]
{\tilde{a}_i^\dagger(\mathbf{m})}
{\tilde{a}}
{_j^\dagger(\mathbf{m'})  \tilde{a}_k^\dagger(\mathbf{l})}
{\tilde{a}}
\contraction[1ex]
{\tilde{a}_i^\dagger(\mathbf{m}) \tilde{a}_j^\dagger(\mathbf{m'})}
{\tilde{a}}
{_k^\dagger(\mathbf{l})
\tilde{a}_k(\mathbf{l})}
{\tilde{a}}
\tilde{a}_i^\dagger(\mathbf{m}) \tilde{a}_j^\dagger(\mathbf{m'})  \tilde{a}_k^\dagger(\mathbf{l})
\tilde{a}_k(\mathbf{l})
\tilde{a}_j(\mathbf{m}') 
\tilde{a}_i(\mathbf{m})    \nonumber\\
\to& -3\times (6g)^2 \left(1+6s_{NL}^2\right)\left[\left(1+6s_{NL}^2\right)^2+17s_{NL}^4\right] \ , 
\end{align}
and
\begin{align}
\sum_{\mathbf{l},\mathbf{m},\mathbf{m'}}&\sum_{ijk}
\contraction[3ex]
{}
{\tilde{a}}
{_i^\dagger(\mathbf{m}) \tilde{a}_j^\dagger(\mathbf{m'})  \tilde{a}_k^\dagger(\mathbf{l})}
{\tilde{a}}
\contraction[2ex]
{\tilde{a}_i^\dagger(\mathbf{m})}
{\tilde{a}}
{_j^\dagger(\mathbf{m'})  \tilde{a}_k^\dagger(\mathbf{l})
\tilde{a}_k(\mathbf{l})
\tilde{a}_j(\mathbf{m}')}
{\tilde{a}}
\contraction[1ex]
{\tilde{a}_i^\dagger(\mathbf{m}) \tilde{a}_j^\dagger(\mathbf{m'})}
{\tilde{a}}
{_k^\dagger(\mathbf{l})
\tilde{a}_k(\mathbf{l})}
{\tilde{a}}
\tilde{a}_i^\dagger(\mathbf{m}) \tilde{a}_j^\dagger(\mathbf{m'})  \tilde{a}_k^\dagger(\mathbf{l})
\tilde{a}_k(\mathbf{l})
\tilde{a}_j(\mathbf{m}') 
\tilde{a}_i(\mathbf{m})    \nonumber\\
\to& \,\, 2\times 6g \Big[
\left(1+6s_{NL}^2\right)^3 
+ 3(s_{NL}^2)^2\left(1+6s_{NL}^2\right) \nonumber\\
&+8(2s_{NL}^2)^2\left(1+6s_{NL}^2\right)
+ 8s_{NL}(2s_{NL}^2)^2\nonumber\\
&+8(2s_{NL}^2)^3 
+ 4 (2s_{NL}^2)^2\left(1+6s_{NL}^2\right)\nonumber\\
&+4 s_{NL}^2(2s_{NL}^2)^2\Big] \nonumber\\
=& \,\, 12g\Big[
\left(1+6s_{NL}^2\right)^3  \nonumber\\
&+51s_{NL}^4\left(1+6s_{NL}^2\right)
+ 112 s_{NL}^6 \Big]     \ .
\end{align}
Here, the prefactors of two and three in the last two contractions are again singled out, and they account for equivalent ways to perform these. In the 
last contraction (having fixed one of the sites, say $\mathbf{m}$) there are 36 different possibilities for $\mathbf{m}'$ and $\mathbf{l}$. They are all accounted for on the right hand side of the arrow, with the individual factors from the anti-commutator~(\ref{comm_mm}) identified in parenthesis. Putting all together we arrive at Eq.~(\ref{Wfull}).

\begin{figure}[t!]
    \centering
    \includegraphics[width=0.49\textwidth]{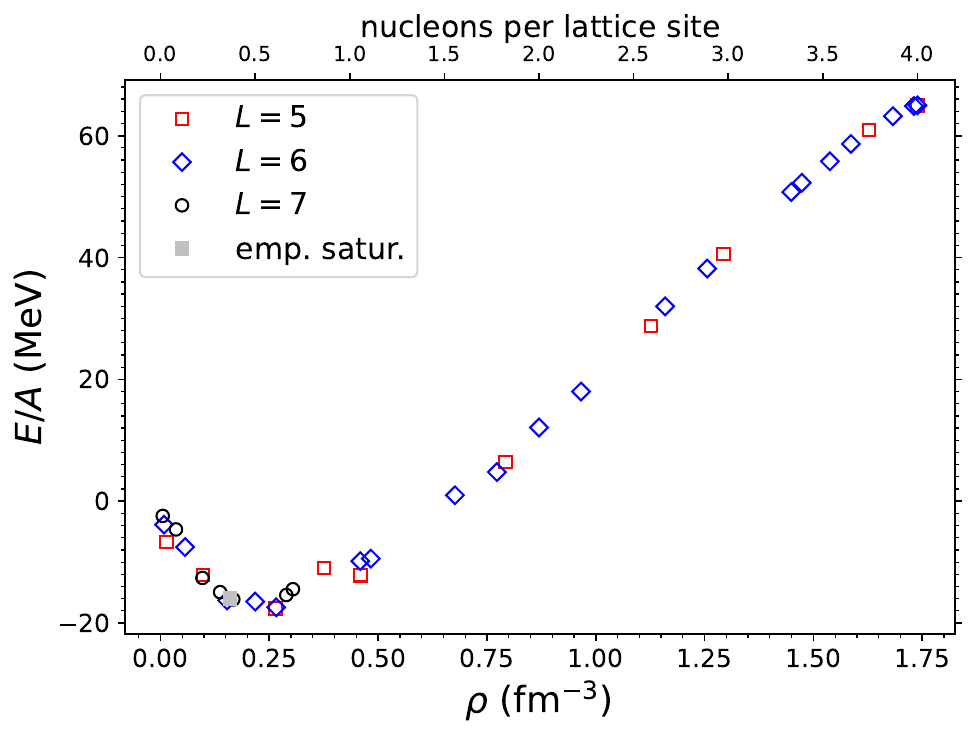}
        \caption{Energy per nucleon $E/A$ in symmetric nuclear matter as a function of density $\rho$ from mean-field expectation values computed on lattices with extent $L$ using the Hamiltonian of Ref.~\cite{lu2019} [See Eq.~(\ref{nopi})]. The  empirical saturation point is shown as a gray rectangle. The top $x$ axis shows the density in nucleons per lattice site.} 
    \label{fig:EOS_2018_full}
\end{figure}

Figure~\ref{fig:EOS_2018_full} shows the equation of state of the Hamiltonian~(\ref{nopi}) from Ref.~\cite{lu2019} over the full range of density. The analytical results match the numerical ones at the maximum density. The region around the saturation point and the individual energy contributions are shown in Figs.~\ref{fig:EOS_2018} and \ref{fig:EOS-contrib} of the main text. 

We turn to the potentials $V_L$ and $V_{NL}$ of the Hamiltonian $H_B$, see Eqs.~(\ref{VL}) and (\ref{eq:VNL}), respectively of Ref.~\cite{elhatisari2016}, and  compute their expectation values on a fully filled lattice. In what follows, it will be useful to occasionally separate the two spin indices $s=1,2$ (which stand for the projections $1/2$ and $-1/2$) and the two similarly defined isospin indices  $i=1,2$  and introduce 
$a_{1,1}(\mathbf{n})=a_1(\mathbf{n})$, $a_{1,2}(\mathbf{n}) = a_2(\mathbf{n})$, $a_{2, 1}(\mathbf{n}) = a_3(\mathbf{n})$, and $a_{2, 2}(\mathbf{n}) = a_4(\mathbf{n})$.

We start with the nonlocal potential $V_{NL}$ of Eq.~(\ref{eq:VNL}). The expectation value from the nonlocal density ${:\rho_{NL}(\mathbf{n})^2:}$ can simply be obtained by setting $s_L=0$ in Eq.~(\ref{full}). 
Thus, we only need to compute the contribution from the isospin density ${:\rho_{I,NL}(\mathbf{n})^2:}$ of Eq.~(\ref{rhoINL}). Because of isospin symmetry of the fully occupied lattice we only need to compute one isospin component and focus on $I=z$. 
We have
\begin{equation}
	\rho_{I=z,NL}(\mathbf{n})=\sum\limits_{s=1}^2\left(\tilde{a}_{s,1}^\dagger(\mathbf{n})\tilde{a}_{s,1}(\mathbf{n}) - \tilde{a}_{s,2}^\dagger(\mathbf{n})\tilde{a}_{s,2}(\mathbf{n})\right) \ .
\end{equation}
Thus, 
\begin{align}
	:{\rho}_{I=z,NL}^2(\mathbf{n}) :=&\sum\limits_{s,s'=1}^2\bigg[\tilde{a}^\dagger_{s,1}(\mathbf{n}) \tilde{a}^\dagger_{s',1}(\mathbf{n}) \tilde{a}_{s',1}(\mathbf{n}) \tilde{a}_{s,1} (\mathbf{n})\notag\\& - \tilde{a}^\dagger_{s,1}(\mathbf{n}) \tilde{a}^\dagger_{s',2}(\mathbf{n}) \tilde{a}_{s',2}(\mathbf{n}) \tilde{a}_{s,1}  (\mathbf{n})\notag\\
	& - \tilde{a}^\dagger_{s,2}(\mathbf{n}) \tilde{a}^\dagger_{s,1}(\mathbf{n}) \tilde{a}_{s',1}(\mathbf{n}) \tilde{a}_{s,2}  (\mathbf{n})\notag\\& + \tilde{a}^\dagger_{s,2}(\mathbf{n}) \tilde{a}^\dagger_{s',2}(\mathbf{n}) \tilde{a}_{s,2}(\mathbf{n}) \tilde{a}_{s',2} (\mathbf{n}) \bigg]\,.
    \label{eq:RhoSquaredNLIsospinZ}
\end{align}
In the fully occupied lattice, the  non-vanishing expectation values come from terms such as 
\begin{align}
	\sum_{s,s'=1}^2 
	\contraction[2ex]{}{\tilde{a}}
	{_{s,1}^\dagger(\mathbf{n})  \tilde{a}_{s',1}^\dagger(\mathbf{n})
		\tilde{a}_{s',1}(\mathbf{n})}{\tilde{a}}
	\contraction[1ex]{\tilde{a}_{s,1}^\dagger(\mathbf{n})}{\tilde{a}}{_{s',1}^\dagger(\mathbf{n})}{\tilde{a}}
	\tilde{a}_{s,1}^\dagger(\mathbf{n})  \tilde{a}_{s',1}^\dagger(\mathbf{n})
	\tilde{a}_{s',1}(\mathbf{n})  \tilde{a}_{s,1}(\mathbf{n}) \to 4\left(1+6s_{NL}^2\right)^2 \ , 
\end{align}
\begin{align}
	\sum_{s,s'=1}^2
	\contraction{}
	{\tilde{a}}
	{_{s,1}^\dagger(\mathbf{n})\tilde{a}_{s',1}(\mathbf{n})}
	{\tilde{a}}
	\contraction[2ex]{\tilde{a}_{s,1}^\dagger(\mathbf{n})}{\tilde{a}}
	{_{s',1}^\dagger(\mathbf{n})
		\tilde{a}_{s',1}(\mathbf{n})}
	{\tilde{a}}
	\tilde{a}_{s,1}^\dagger(\mathbf{n})  \tilde{a}_{s',1}^\dagger(\mathbf{n})
	\tilde{a}_{s',1}(\mathbf{n})  \tilde{a}_{s,1}(\mathbf{n}) \to -2\left(1+6s_{NL}^2\right)^2 \ , 
\end{align}
and
\begin{align}
	\sum_{s,s'=1}^2 
	\contraction[2ex]{}{\tilde{a}}
	{_{s,1}^\dagger(\mathbf{n})  \tilde{a}_{s',2}^\dagger(\mathbf{n})
		\tilde{a}_{s',2}(\mathbf{n})}{\tilde{a}}
	\contraction[1ex]{\tilde{a}_{s,1}^\dagger(\mathbf{n})}{\tilde{a}}{_{s',2}^\dagger(\mathbf{n})}{\tilde{a}}
	\tilde{a}_{s,1}^\dagger(\mathbf{n})  \tilde{a}_{s',2}^\dagger(\mathbf{n})
	\tilde{a}_{s',2}(\mathbf{n})  \tilde{a}_{s,1}(\mathbf{n}) \to 4\left(1+6s_{NL}^2\right)^2 \ .
\end{align}
There are three more analogous contractions from the final two terms in Eq.~\eqref{eq:RhoSquaredNLIsospinZ}.
The sum gives 
\begin{equation}
	\langle:\rho_{I=z,NL}^2(\mathbf{n}) :\rangle = -4(1+6s_{NL}^2)^2\,.
\end{equation}
This is just for $I=z$, so we get a factor of 3 from all three isospin $I$ components. Finally, dividing by 4 nucleons per lattice site and accounting for the prefactors in Eq.~\eqref{eq:VNL}, 
the complete result is
\begin{equation}
\label{expVNL}
    \frac{\langle V_{NL}\rangle}{A} = \frac{3}{2}\left(c_{NL}-c_{I,NL}\right)\left(1+6s_{NL}^2\right)^2\ .
\end{equation}

We turn to the local potential $V_L$ of Eq.~(\ref{VL}). The expectation value of ${:\rho_L(\mathbf{n})^2:}$ can be obtained from Eq.~(\ref{full}) by setting $s_{NL}=0$. For the remaining spin, isospin, and spin-isospin terms we note that we only need to compute one of them because of the spin-isospin symmetry of the fully occupied lattice, and we can again focus on  a single component. 

\begin{widetext}
Let us take $\rho_{S=x,I=x}$ for example. 
Then, 
\begin{align}
	\rho_{S=x,I=x}(\mathbf{n}) &= a^\dagger(\mathbf{n})\left[\sigma_x\otimes\tau_x\right]a(\mathbf{n})\\
	&=a^\dagger_1(\mathbf{n})a_4(\mathbf{n}) - a_2^\dagger(\mathbf{n}) a_3 (\mathbf{n})- a^\dagger_3(\mathbf{n})a_2(\mathbf{n})+a^\dagger_4(\mathbf{n})a_1(\mathbf{n})\notag\\
	&+s_L\sum_{|\mathbf{n}-\mathbf{m}|=1}\Big[a^\dagger_1(\mathbf{n})a_4(\mathbf{m}) - a_2^\dagger(\mathbf{n}) a_3 (\mathbf{m})- a^\dagger_3(\mathbf{n})a_2(\mathbf{m})+a^\dagger_4(\mathbf{n})a_1(\mathbf{m})\Big]\ .
\end{align}
When computing $:\rho^2_{S=x,I=x}(\mathbf{n}):$ one gets a lot of terms. However, as we need to match spin-isospin indices when computing the Wick contractions in the fully occupied lattice, only a smaller number of terms survive.  These are
\begin{align}
\label{temp}
	:\rho^2_{S=x,I=x}(\mathbf{n}): &\to  
    a^\dagger_1(\mathbf{n})a^\dagger_4(\mathbf{n})a_1(\mathbf{n})a_4(\mathbf{n})
    +a^\dagger_2(\mathbf{n})a^\dagger_3(\mathbf{n})a_2(\mathbf{n})a_3(\mathbf{n})\notag\\
	&+2s_L\sum\limits_{|\mathbf{n}-\mathbf{m}|=1}\Big[
    a^\dagger_1(\mathbf{n})a^\dagger_4(\mathbf{m})a_1(\mathbf{m})a_4(\mathbf{n})
    +a^\dagger_2(\mathbf{n})a^\dagger_3(\mathbf{m})a_2(\mathbf{m})a_3(\mathbf{n})
    \Big] \nonumber \\
	&+s_L^2\sum_{\substack{|\mathbf{n}-\mathbf{m}|=1\\|\mathbf{n}-\mathbf{m'}|=1}}
    \Big[
    a^\dagger_1(\mathbf{m'})a^\dagger_4(\mathbf{m})a_1(\mathbf{m})a_4(\mathbf{m'})
	+a^\dagger_2(\mathbf{m'})a^\dagger_3(\mathbf{m})a_2(\mathbf{m})a_3(\mathbf{m'})
    \Big] +\mathrm{h.c.}\,.
\end{align}
\end{widetext}

The non-vanishing Wick contractions yield 
\begin{align}
	\contraction{}
	{{a}}
	{_{1}^\dagger(\mathbf{n}){a}_{1}(\mathbf{n})}
	{\tilde{a}}
	\contraction[2ex]{{a}_{4}^\dagger(\mathbf{n})}{{a}}
	{_{1}^\dagger(\mathbf{n})
		{a}_{4}(\mathbf{n})}
	{{a}}
	{a}_{1}^\dagger(\mathbf{n})  {a}_{4}^\dagger(\mathbf{n})
	{a}_{1}(\mathbf{n})  {a}_{4}(\mathbf{n}) \to -1 \ , 
\end{align}
 and 
\begin{align}
	\sum_{\substack{|\mathbf{n}-\mathbf{m}|=1\\|\mathbf{n}-\mathbf{m'}|=1}}&
	\contraction{}
	{{a}}
	{_{1}^\dagger(\mathbf{m'}){a}_{1}(\mathbf{m})}
	{{a}}
	\contraction[2ex]{{a}_{4}^\dagger(\mathbf{m})}{{a}}
	{_{1}^\dagger(\mathbf{m})
		{a}_{4}(\mathbf{m'})}
	{{a}}
	{a}_{1}^\dagger(\mathbf{m'})  {a}_{4}^\dagger(\mathbf{m})
	{a}_{1}(\mathbf{m})  
    {a}_{4}(\mathbf{m'}) \to -6 \ . 
\end{align}

Putting this all together, we get
\begin{align}
	\langle:\rho_{S=x,I=x}^2(\mathbf{n}):\rangle =& -4(1+6s_L^2) \ , 
\end{align}
and the factor of 4 comes from the Hermitian conjugates and $(1,4)\to(2,3)$ in Eq.~(\ref{temp}).

The final result is
\begin{align}
    \frac{\langle V_L\rangle}{A} &= \frac{3}{2}c_L\left(1+16s_L+46s_L^2\right)\notag\\
    &-\frac{3}{2}\left(c_{S,L}+c_{I,L}+3c_{S,I,L}\right)\left(1+6s_L^2\right)\ .
\end{align}

\twocolumngrid

\section{Expectation values at lowest densities}
The lowest densities we compute are at $\rho=4/(aL)^3$, where the zero-momentum state of four nucleons is
\begin{align}
    |\mathbf{p}=0\rangle &\equiv L^{-6} \sum_{\mathbf{k},\mathbf{l},\mathbf{m},\mathbf{n}} a_1^\dagger(\mathbf{k}) a_2^\dagger(\mathbf{l}) a_3^\dagger(\mathbf{m}) a_4^\dagger(\mathbf{n}) |0\rangle \nonumber\\
    &=L^{-6}\sum_{\mathbf{k},\mathbf{l},\mathbf{m},\mathbf{n}} \frac{\tilde{a}_1^\dagger(\mathbf{k}) \tilde{a}_2^\dagger(\mathbf{l}) \tilde{a}_3^\dagger(\mathbf{m}) \tilde{a}_4^\dagger(\mathbf{n})}{\left(1+6s_{NL}\right)^4} |0\rangle \ .
\end{align}

Here, the translational invariance of the zero-momentum state allowed us to rewrite it in terms of the $\tilde{a}^\dagger_i$ operators. It is now straightforward (with all we learned in Sec.~\ref{full_derivation}) to compute the expectation value (using $A=4$)
\begin{align}
    \frac{\langle \mathbf{p}=0|V_0|\mathbf{p}=0 \rangle}{A} &= \frac{3c_0}{2L^3} \left(1+6s_{L}\right)^2\left(1+6s_{NL}\right)^4 \nonumber\\
    &= \frac{3}{8}c_0 a^{3} \left(1+6s_{L}\right)^2\left(1+6s_{NL}\right)^4\rho \ .
\end{align}
This is Eq.~(\ref{lowdens}). The expression is intuitively clear. The nonlocal and local smearing each involves six sites. Nonlocal smearing involves four annihilation and creation operators (hence the power four), while local smearing involves two densities. 

\section{Energies from the transfer matrix}
The auxiliary field Monte Carlo calculations of Refs.~\cite{epelbaum2010,epelbaum2010b,epelbaum2011,epelbaum2012,epelbaum2013,epelbaum2014,elhatisari2016,elhatisari2017} solved the many-body problem by applying the transfer matrix formalism~\cite{lee2009} with a discrete temporal lattice spacing $a_t=1/(150\,\mathrm{MeV})$. In a nutshell, this works as follows.

One introduces the transfer matrix
\begin{equation}
M(\phi) = : e^{-a_t H_1(\phi)} : 
\end{equation}
Here $H_1$ is a one-body Hamiltonian that depends on auxiliary fields $\phi$ after a Hubbard Stratonovich transformation of original Hamiltonian $H$ (that included two- and three-body interactions). We note the occurrence of the temporal lattice spacing $a_t$. Next, one introduces a partition function via the path integral
\begin{equation}
\label{Z}
    Z=\int {\cal D}\phi 
    e^{-S(\phi)} \langle\psi|M_{L_t-1}(\phi) \cdots M_0(\phi)|\psi\rangle \ , 
\end{equation}
where $|\psi\rangle$ is an initial state and $L_t$ temporal steps are being made. (The subscripts on the transfer matrices help us counting the number $L_t$ of them.) The idea is that the ``time evolution'' $L_ta_t\to\infty$ projects out the ground state. Then the energy is computed as
\begin{equation}
\label{EfromZ}
    E = \lim_{L_t\to\infty} \frac{\log Z}{a_t L_t} \ .
\end{equation}
Strictly speaking, however, one needs $L_ta_t\to\infty$ while keeping $a_t$ sufficiently small. (Sufficiently small means that $a_t E \ll 1$.)  However, Refs.~\cite{epelbaum2010b,epelbaum2011,epelbaum2012,epelbaum2013,epelbaum2014,elhatisari2016,elhatisari2017} used $a_t=1/(150~\mathrm{MeV})$.

Assuming that $a_t$ is sufficiently small, one can rewrite
the partition function~(\ref{Z}) as 
\begin{equation}
    Z\approx\int {\cal D}\phi 
    e^{-S(\phi)} \langle\psi|1-L_t a_t H_1|\psi\rangle \ ,
\end{equation}
and it is clear that the correct energy is obtained from Eq.~(\ref{EfromZ}). 

In contrast, if $a_t$ is not sufficiently small then the computed energy
\begin{equation}
\label{simple}
    E_{a_t} = -\frac{\log(1 - a_t E_\mathrm{true})}{a_t}
\end{equation}
is not necessarily close to the true energy $E_\mathrm{true}$.
Indeed, using the Hartree-Fock energies in Table~\ref{tab:res_combo} as the true energies and $a_t=1/(150~\mathrm{MeV})$, one does get the corresponding results from auxiliary field Monte Carlo to a good approximation.

One can also turn this around. Using the temporal spacing $a_t=1/(150~\mathrm{MeV})$ and Eq.~(\ref{EfromZ}) one does not really compute the energy of the original Hamiltonian $H$ but rather that of a different Hamiltonian containing various powers of $H$. We finally note that the temporal lattice spacing was decreased to $a_t=1/(1000~\mathrm{MeV})$ in Ref.~\cite{lu2019}, and the energy was not computed any more from Eq.~(\ref{EfromZ}) but rather from an expectation value of the Hamiltonian, see, e.g., Refs.~\cite{elhatisari2024,niu2025}.  
\end{document}